\documentclass{jfm}

\usepackage{graphicx}
\graphicspath{{Figures/}}
\usepackage{amsmath,amssymb,amsfonts}
\usepackage{subfigure}
\usepackage{multirow}
\DeclareMathOperator*{\argmax}{argmax}

\begin{document}

\newtheorem{lemma}{Lemma}
\newtheorem{corollary}{Corollary}

\shorttitle{Sub-grid scale model classification and blending through deep learning} 
\shortauthor{R. Maulik et al} 

\title{Sub-grid scale model classification and blending through deep learning}

\author
 {
 Romit Maulik\aff{1}
 \and 
 Omer San\aff{1}
 \corresp{\email{osan@okstate.edu}},
 \and
 Jamey D. Jacob\aff{1}
 \and
 Christopher Crick\aff{2}
 }

\affiliation
{
\aff{1}
Mechanical \& Aerospace Engineering, Oklahoma State University, Stillwater, OK 74078, USA
\aff{2}
Computer Science, Oklahoma State University, Stillwater, OK 74078, USA
}

\maketitle

\begin{abstract}
In this article we detail the use of machine learning for spatiotemporally dynamic turbulence model classification and hybridization for the large eddy simulations (LES) of turbulence. Our predictive framework is devised around the determination of local conditional probabilities for turbulence models that have varying underlying hypotheses. As a first deployment of this learning, we classify a point on our computational grid as that which requires the functional hypothesis, the structural hypothesis or no modeling at all. This ensures that the appropriate model is specified from \emph{a priori} knowledge and an efficient balance of model characteristics is obtained in a particular flow computation. In addition, we also utilize the conditional probability predictions of the same machine learning to blend turbulence models for another hybrid closure. Our test-case for the demonstration of this concept is given by Kraichnan turbulence which exhibits a strong interplay of enstrophy and energy cascades in the wave number domain. Our results indicate that the proposed methods lead to robust and stable closure and may potentially be used to combine the strengths of various models for complex flow phenomena prediction.
\end{abstract}

\section{Introduction}

Turbulence is an active area of research due to its significant impact on a diverse set of challenges such as those pertaining to the aerospace and geophysical communities. In recent decades, computational fluid dynamics (CFD) has proven to be useful for low-cost realizations of flow phenomena for critical decision making processes. However, CFD is still fairly limited in terms of accuracy due to the exceptional computational expense involved in high-fidelity simulations of turbulence. `True' numerical experiments require the use of a direct numerical simulation (DNS) of the Navier-Stokes equations. DNS is only possible if a discretized domain can resolve all possible frequencies in a flow and is therefore out of reach of the vast majority of engineering and geophysical applications for the foreseeable future. Large eddy simulations (LES) \citep{sagaut2006large} have proven to be a promising strategy for resolving a greater number of scales in a flow but require the specification of a model which represents the interactions of the higher frequencies with the mean flow. This sub-grid scale (SGS) model, also known as a closure, is usually specified in the form of an algebraic or differential equation and is generally flow category specific \citep{vreman2004eddy}. This imposes a caveat on the applicability of a SGS model if no \emph{a priori} information of the flow category is known. As such, the basic premise of LES is extendable to many partial differential equation systems with quadratic non-linearities. In this paper, we explore the utilization of machine learning for dynamically inferring regions where a particular turbulence modeling hypothesis is applicable with the goal of improving predictive capabilities of turbulence dynamics for a wide range of problems.

The multi-scale nature of turbulence requires the use multiple modeling approximations for the higher wavenumbers which remain unsupported by computational degrees of freedom (a case for most flows of any practical interest).  The procedure of modeling these smaller scales is often denoted \emph{closure} due to insufficient knowledge about higher-order wavenumber interactions with the coarse-grained system \citep{berselli2006mathematics} and remains vital for the accurate computation of many applications \citep{hickel2014subgrid,yu2016dynamic}. From an LES point of view, the closure problem may be considered to be dominated by commutative errors in the calculation of the non-linear term as well as the defects associated with commutative errors stemming from the dynamic term. In this study, we focus on the former.

Explicit LES argues for the utilization of sub-grid models specified as algebraic or differential equations for the unresolved scales. These are built on an intuitive reasoning of how the losses of coarse-graining the Navier-Stokes equations may be incorporated into an LES deployment. Some of the most notable sub-grid closure strategies are those given by the linear eddy-viscosity hypothesis, which models the sub-grid stress tensor through the Bousinessq approximation. Within the context of the Navier-Stokes equations, it is generally accepted that the vorticity dominated smaller scales are dissipative \citep{kolmogorov1941local} and therefore, most turbulence models seek to specify a sub-grid dissipation \citep{frisch1995turbulence}. Many functional sub-grid models can be traced back to \cite{smagorinsky1963general}, where an effective eddy-viscosity was determined by an \emph{a priori} specified mixing length and a $k^{-5/3}$ scaling recovery for the kinetic energy content in the wavenumber domain. Similar hypotheses have also been used for two-dimensional turbulence (often utilized as a test-bed for geophysical scenarios, for instance see \cite{mcwilliams1990vortices,tabeling2002two,boffetta2012two,pearson2017evaluation,pearson2018log}), for approximating the $k^{-3}$ cascade and generally have their roots in dimensional analysis related to the cascade of enstrophy \citep{leith1968diffusion}. These models may also be classified as \emph{functional} due to the phenomenological nature of their deployment and comprise the bulk of explicit LES turbulence models used in practical deployments. Explicit LES closures may also be specified through the specification of a low-pass spatial filter to account for the unresolved scales \citep{bardina1980improved,stolz1999approximate,layton2003simple,mathew2003explicit,san2018generalized} where phenomenology is bypassed but ansatz are provided for the bulk dissipative nature of the smaller scales through the control of a characteristic filter-width. In either scenario, (i.e., whether structural or functional), the choice of the phenomenology (or dissipation control parameter) plays a key role in the successful calculation of accurate \emph{a posteriori} statistics. 

The past few years have seen a rapid increase in the use of machine learning for various scientificand engineering applications. For turbulence, some widely used strategies for prediction and inference include symbolic regression such as in \cite{weatheritt2016novel,weatheritt2017development,weatheritt2017hybrid}, where functional model-forms for Reynolds-averaged Navier-Stokes (RANS) deployments were generated through evolutionary optimization against high-fidelity data. Other techniques incorporating Bayesian ideologies have also been used, for instance in \cite{xiao2016quantifying} where an iterative ensemble Kalman method was used to assimilate prior data for quantifying model form uncertainty in RANS models. In \cite{wang2017physics,wang2017comprehensive} and \cite{wu2018data}, random-forest regressors were utilized for RANS turbulence-modeling given direct numerical simulation (DNS) data. In \cite{singh2016using} and \cite{singh2017machine}, an ANN was utilized to predict a non-dimensional correction factor in the Spalart-Allmaras turbulence model through a field-inversion process using experimental data. Bypassing functional formulations of a turbulence model was also studied from the RANS point of view by \cite{tracey2015machine}. \cite{ling2015evaluation} utilized support vector machines, decision trees and random forest regressors for identifying regions of high RANS uncertainty. A deep-learning framework where Reynolds-stresses would be predicted in an invariant subspace was developed by \cite{ling2016reynolds}. Machine learning of invariance properties has also been discussed in the context of turbulence modeling by \cite{ling2016machine}. The reader is directed to a recent review by \cite{duraisamy2018turbulence}, for an excellent review of turbulence modeling using data-driven ideas.

As shown above, the use of data-driven ideologies and in particular artificial neural networks (ANNs) has generated significant interest in the turbulence modeling community for addressing long-standing challenges (also see \cite{sotgiu2018turbulent,zhu2019machine,zhang2019application} for recent examples). A multilayered ANN may be optimally trained to approximate any non-linear function \citep{hornik1989multilayer} and the large data sets involved in turbulence research couple with ever-improving computing capabilities has also motivated the study of ANN based learning. Within the context of LES (and associated with the scope of this paper) there are several investigations into sub-grid modeling using data-driven techniques. In an early study of the feasibility of using learning from DNS, \cite{sarghini2003neural} deployed ANNs for estimating Smagorinsky model-form coefficients within a mixed sub-grid model for a turbulent channel flow. ANNs were also used for wall-modeling by \cite{milano2002neural} where it was used to reconstruct the near wall field and compared to standard proper-orthogonal-decomposition techniques. An alternative to ANNs for sub-grid predictions was proposed by \cite{king2016autonomic} where \emph{a priori} optimization was utilized to minimize the $L^2$-error between true and modeled sub-grid quantities using a parameter-free Volterra series. \cite{maulik2017neural} utilized an extreme-learning-machine (a variant of a single-layered ANN) to obtain maps between low-pass spatially filtered and deconvolved variables in an \emph{a priori} sense. This had implications for the use of ANNs for turbulence modeling without model-form specification. A more in-depth investigation was recently undertaken by \cite{fukami2018super} where convolutional ANNs were utilized for reconstructing from downsampled snapshots of turbulence. \cite{maulik2018deconvolution} also deployed a data-driven convolutional and deconvolutional operation to obtain closure terms for two-dimensional turbulence. \cite{gamahara2017searching} utilized ANNs for identifying correlations with grid-resolved quantities for an indirect method of model-form identification in turbulent channel flow. The study by \cite{vollant2017subgrid} utilized ANNs in conjuction with optimal estimator theory to obtain functional forms for sub-grid stresses. In \cite{beck2018neural}, a variety of neural network architectures such as convolutional and recurrent neural networks are studied for predicting closure terms for decaying homogeneous isotropic turbulence. A least-squares based truncation is specified for stable deployments of their model-free closures. Model-free turbulence closures are also specified by \cite{maulik2018deconvolution,maulik2019subgrid} and \cite{wang2018investigations}, where sub-grid scale stresses are learned directly from DNS data and deployed in \emph{a posteriori} assessments. \cite{king2018deep} studied generative-adversarial networks and the LAT-NET \cite{hennigh2017lat} for \emph{a priori} recovery of statistics such as the intermittency of turbulent fluctuations and spectral scaling. 

While a large majority of the LES-based frameworks presented above utilize a least-squares error minimization technique for constructing maps to sub-grid stresses \emph{directly} for theoretically optimal LES \citep{langford1999optimal,moser2009theoretically,labryer2015framework}, this work is novel in that it utilizes sub-grid statistics (pre-computed from DNS data) to train a classifier.
Our trained intelligence utilizes the most appropriate turbulence modeling modelling hypothesis (i.e., either structural or functional) from \emph{a priori} experience to close the LES governing equations. It is also deployed to blend turbulence models linearly at each point during flow evolution for a novel hybrid closure. In this manner, we are able to co-deploy models having fundamentally different underlying hypothesis for turbulence parameterizations in a stable manner. This is similar to the study employed in \cite{ling2017data} where machine learning is utilized for adaptive error corrections in RANS deployments. In the rest of this article, we discuss the governing equations of decaying Kraichnan turbulence, introduce our machine learning architecture and its optimization and detail its \emph{a priori} and \emph{a posteriori} performance through statistical assessments.

\section{Governing equations}

We proceed by outlining our Kraichnan turbulence test-case which (alongwith quasigeostrophic turbulence) is an important prototype for geophysical flow-phenomenon with high aspect ratios and for which turbulence model research remains highly active \citep{pearson2017evaluation}. The governing equations of motion for Kraichnan turbulence are given by the two-dimensional Navier-Stokes equations in a periodic domain. The non-dimensionalized version of these equations may be expressed in the vorticity ($\omega$) and stream function ($\psi$) formulation as \citep{san2012high}, 
\begin{align}
\begin{gathered}
\frac{\partial \omega}{\partial t} + J(\omega,\psi) = \frac{1}{Re} \nabla^2 \omega, \\
x, y \in [0,2 \pi], t \in [0,4],
\end{gathered}
\end{align}
where we define the Jacobian (or the nonlinear term as)
\begin{align}
J(\omega,\psi) = \frac{\partial \omega}{\partial x} \frac{\partial \psi}{\partial y} - \frac{\partial \omega}{\partial y} \frac{\partial \psi}{\partial x},
\end{align}
and the conservation of mass is enforced by 
\begin{align}
\nabla^2 \psi = -\omega.
\end{align}
A measure of multi-scale behavior in this system is given by the Reynolds number ($Re$). A high value of $Re$ combined with a coarse-grid projection of these equations results in insufficient support for the finest structures in the flow evolution, leading to noise accumulation at grid cut-off and potential floating point overflow of the numerical evolution of this problem. A sufficiently coarse-grained representation of the governing equations introduced previously are given by the LES governing equations
\begin{align}
\begin{split}
\frac{\partial \bar{\omega}}{\partial t} + J(\bar{\omega},\bar{\psi}) &= \frac{1}{Re} \nabla^2 \bar{\omega} + \Pi, \\
\nabla^2 \bar{\psi} &= -\bar{\omega},
\end{split}
\end{align}
where $\Pi$ may be assumed to be the perfect closure given by 
\begin{align}
\Pi = J(\bar{\omega},\bar{\psi}) - \overline{J(\omega,\psi)}.
\end{align}
When adequately simulated, the Kraichnan turbulence test cases results in the classical $k^{-3}$ scaling of the energy spectra \citep{kraichnan1967inertial}. In practice, this perfect estimation of loss is never available in a numerical deployment and must be estimated by either an algebraic or differential equation. We focus on two competing ideologies for estimating closure. The first is given by the functional hypothesis and may be expressed as
\begin{align}
\Pi = \nu_e \nabla^2 \bar{\omega}
\end{align}
where the Smagorinsky approximation to the eddy-viscosity $\nu_e$ is given by
\begin{align}
\begin{gathered}
\nu_e = (C_s \delta)^2 |\bar{S}|, \\
|\bar{S}| = \sqrt{4 \Big(\frac{\partial^2 \bar{\psi}}{\partial x \partial y}\Big)^2 + \Big(\frac{\partial^2 \bar{\psi}}{\partial x^2} - \frac{\partial^2 \bar{\psi}}{\partial y^2}\Big)^2},
\end{gathered}
\end{align}
where it is very common to consider the filter length scale $\delta$ as the representative mesh size. 
A successful application of this closure necessitates a dynamic calculation of the Smagorinsky coefficient $C_s$ that requires the specification of a test-filter and a spatial-averaging for stabilized deployment. This approach is the well-known dynamic Smagorinsky (DS) closure \citep{germano1991dynamic} and its two-dimensional abstraction for Kraichnan turbulence has been presented by \cite{san2014dynamic}.

A competing ideology is given by the structural (or scale-similarity) hypothesis which assumes that the LES equations are projections of the Navier-Stokes equations to a smoother space where an inverse-filtering operator may be utilized to recover the finer scales that are lost. Mathematically,
\begin{align}
\Pi = J(\bar{\omega},\bar{\psi}) - \widetilde{J(\omega^*, \psi^*)}
\end{align}
where $\omega^*$ and $\psi^*$ are approximately deconvolved variables obtained through an inverse-filtering procedure and a Gaussian-type filter kernel (given by the $G(a)=\tilde{a}$) is an approximation of the projection to the LES space. However, these techniques are limited due to the underlying assumption of isomorphism between the LES and the Navier-Stokes equations \citep{germano2015similarity}. In practice, this implies that structural hypotheses are appropriate only if finer structures are sufficiently well-resolved on a particular grid. As such, this diminishes their benefit for practical flows where grid cut-off wave numbers are generally much smaller than the largest wave number in the flow. The breakdown of structural closures manifests itself in the form of stability issues. For this reason, many successful closure deployments utilize linear combinations of structural and functional models \citep{habisreutinger2007coupled}. In this work, we implement approximate-deconvolution (AD) \citep{stolz1999approximate} which utilizes an iterative application of the trapezoidal filter kernel for inversion of filtered grid-quantities and utilize three iterative resubstitutions to deconvolve our grid-resolved variables. 

\begin{figure}
\centering
\includegraphics[width=0.9\textwidth]{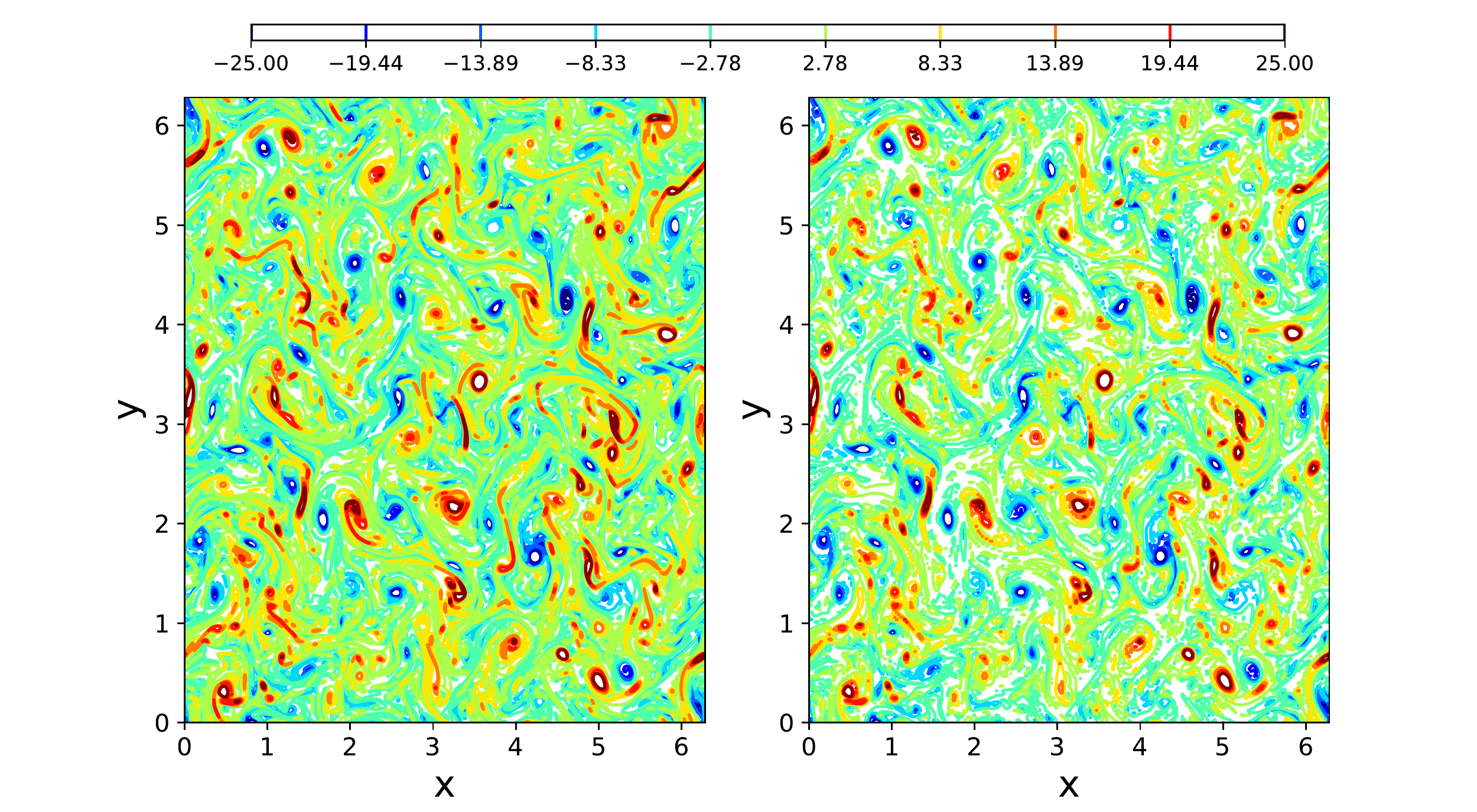}
\caption{Visualization of the effect of Fourier cut-off filtering with DNS (left) and corresponding FDNS (right).}
\label{Fig1}
\end{figure}

\section{Machine learning}

\begin{figure}
\centering
\includegraphics[trim={3cm 19cm 6cm 1cm},clip,width=0.8\textwidth]{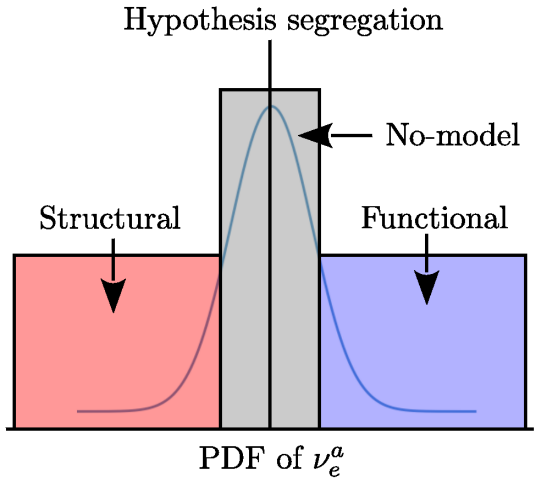}
\caption{Data-segregation for one-hot labeling. The \emph{a priori} eddy-viscosities are projected onto a Gaussian distribution where data beyond 1.0\% of the standard-deviation is labeled as requiring structural (if negative) or functional (if positive) modeling. The remaining data points are classified as no-model cases.}
\label{Fig2}
\end{figure}

We now turn to procedure of utilizing DNS data for learning when to switch between one of three closure scenarios. Of these three options, two are given by the choice of the functional hypothesis and AD. The third option is that of a no-model scenario where our learning determines that closure-modeling is unnecessary. The third scenario is retained since there is a possibility for localized areas in a flow resolution to have adequate grid-support so that the contributions of the sub-grid scale become negligible. This switching between scenarios is spatio-temporally dynamic. Before proceeding, we note that the functional deployment eschews the dynamic procedure and simply sets a $C_s=1.0$ for the calculation of the eddy-viscosity as utilized in the standard Smagorinsky model \citep{smagorinsky1963general} as is common in geophysical scenarios \citep{cushman2011introduction}. This ensures that the proposed framework remains computationally tractable, easily interpretable and deployable. We proceed by outlining our learning strategy through the utilization of DNS data. Five snapshots of DNS data at $Re=32000$ and at $2048^2$ degrees of freedom (from 40000 available snapshots) are utilized to compute the grid-filtered variables (herein denoted by FDNS) at $256^2$ degrees of freedom through the application of a spectral cut-off filter. Perfect closure values ($\Pi$) are then obtained (the reader is directed to \cite{maulik2019subgrid} for details related to the calculation of these quantities). Figure \ref{Fig1} visually quantifies the effect of the spectral domain filtering where the FDNS of a snapshot of vorticity is shown. We then introduce the \emph{a priori} eddy-viscosity given by
\begin{align}
\nu_e^a = \frac{\Pi}{\nabla^2 \bar{\omega}}
\end{align}
where all the terms on the right-hand side of the above equation are available through calculation from the DNS snapshots. The \emph{a priori} eddy viscosity is centered at a value of zero (corresponding to a region where closure modeling is unnecessary) and has tails in the negative and positive directions (a hallmark of isotropic turbulence). A core component of the hypothesis in this work stems from the fact that structural hypotheses are not limited to positive eddy-viscosity predictions alone. The reader may note that models utilizing the functional hypothesis always lead to positive eddy-viscosities. The \emph{a priori} eddy viscosities calculated from the DNS data are then projected onto a Gaussian distribution where values lying at a distance of 1\% of the standard-deviation from the mean (of zero) are labeled as those requiring no closure (due to the low strength of the \emph{a priori} eddy-viscosity). Values lying beyond this range are labeled as functional or structural, depending on if they are positive or negative, respectively. This information is encoded in one-hot labeling for a classification deployment and a corresponding schematic for this hypothesis segregation is shown in Figure \ref{Fig2}. It is observed that the a large portion of the available data lies within the first standard deviation of the mean eddy-viscosity. This leads to the potential of turbulence modeling classification being considered from outlier detection point-of-view. A factor which motivates the choice of the Gaussian distribution is the nature of Kraichnan turbulence (which is isotropic in nature with Gaussian statistics). However, we note that machine learning algorithms are also capable of classifying data belonging to complex distributions and that this hypothesis segregation may be tuned for better accuracy. Also, the choice of the 1\% hyper-parameter is also motivated by observing \emph{a posteriori} training accuracies where it is noticed that a relatively simple architecture (mentioned next) is efficiently able to discern the varying hypothesis. Values greater than 1\% for model delineation led to reduced learning accuracies indicating that a physical delineation potentially exists in this projection and categorization. Further study for adding complexity to the hypothesis segregation is necessary.

Each label for the \emph{a priori} eddy-viscosity is also associated with an input kernel of grid-resolved quantities. This kernel is given by a stencil of 9 inputs of vorticity and stream function each (for a total of 18 input variables). These 9 inputs of each field are given by a query of the field quantity at a point on the coarse grid, the 4 adjacent points in each dimension and the 4 diagonally adjacent points. Each sample of our training data thus consists of 18 inputs of vorticity and stream function and outputs given by one-hot labels for the choice of closure modeling strategy. In this article, we have leveraged the fact that the mean of vorticity and streamfunction are both very close to zero and do not necessitate normalization. In addition, the non-dimensionalized formulation of the governing equations implies that our inputs are all dimensionless. However, we note that for practical deployments of any local-kernel based machine learning queries, grid-resolved quantities must be normalized and non-dimensionalized. 

Mathematically, if our input vector $\textbf{p}$ resides in a $P$-dimensional space and our desired output $\textbf{q}$ resides in a $Q$-dimensional space, this framework establishes a map $\mathbb{M}$ as follows:
\begin{align}
\label{eq6}
\mathbb{M} : \{ p_1, p_2, \hdots, p_P\} \in \mathbb{R}^P \rightarrow \{ q_1, q_2, \hdots, q_Q\} \in \mathbb{R}^Q.
\end{align}
Accordingly, the framework utilized in this article leads to the following relation:
\begin{align}
\label{eq7}
\begin{gathered}
\mathbb{M} : \{ \textbf{p} \} \in \mathbb{R}^{18}  \rightarrow \{ P(\textbf{q}|\textbf{p})\} \in \mathbb{R}^3,
\end{gathered}
\end{align}
where our input and output spaces are given by
\begin{align}
\label{eq8}
\begin{gathered}
\textbf{p}_{i,j} = \{ \bar{\omega}_{i,j}, \bar{\omega}_{i,j+1}, \bar{\omega}_{i,j-1}, \hdots, \bar{\omega}_{i-1,j-1}, \bar{\psi}_{i,j}, \bar{\psi}_{i,j+1}, \bar{\psi}_{i,j-1}, \hdots, \bar{\psi}_{i-1,j-1} \} , \\
\textbf{q}_{i,j} = \{ P(\Pi^k_{i,j}| \textbf{p}_{i,j})\},
\end{gathered}
\end{align}
where $i,j$ refer to the spatial indices on the coarse-grid (i.e., the point of deployment) and $k$ refers to the choice of closure scenario (i.e., structural, functional or no closure). We note here that the choice of the local stencil for ANN query reflects the discretization of the governing equations (with second-order accurate stencils requiring a $\pm 1$ query) and the use of the trapezoidal filter in AD. Also, note that our choice of input space is given by raw variable queries rather than derivatives (or other such engineered terms). This is motivated by an aversion to specify bias towards any particular quantity that may otherwise by learned implicitly by the network. However, we note that the classification workflow may benefit significantly from the inclusion of a feature engineering step prior to optimization. This is a subject of ongoing investigation.

Our optimal map $\mathbb{M}$ is then trained by the following loss-function
\begin{align}
E(\textbf{w}) = -\sum_{n=1}^{N} \sum_{k=1}^{K} \{ t_{nk} \log(y_{nk}) + (1-t_{nk})\log(1-y_{nk})\},
\end{align}
where $\textbf{w}$ are the tunable weights and biases of the network, $N$ is the total number of samples and $K=3$ is the total number of closure scenarios. Here, $t_{nk}$ refers to the target (or true) label of class $k$ and sample $n$ and $y_{nk}$ refers to its corresponding prediction. Note that one-hot encoding ensures that $t_{nk}$ values are always binary \cite{Bishop:2006:PRM:1162264}. For reference, our architecture is trained using the open-source deep learning software Tensorflow and is optimized with the use of ADAM, a popular gradient-descent based optimizer.

Our learning architecture is given by a 5 hidden-layer deep neural network with 40 neurons each for calculating the conditional probabilities of the three closure scenarios pointwise in space and time. The hidden layer neurons employ a rectified-linear activation and the output-layer gives us softmax probabilities for the three classes. The scenario with the highest conditional probability is then deployed for model classification i.e.,
\begin{align}
\Pi^{ML}_{i,j} = \Pi^{k}_{i,j} \text{ s.t. } \argmax_k P(\Pi^{k}_{i,j} | \textbf{p}_{i,j})
\end{align}
where $\Pi^{ML}_{i,j}$ refers to the machine learning based turbulence model computation at a point. In the case of model blending, the conditional probabilities for closure scenarios are used to find a linear combination of the standard Smagorinsky and the AD closures. In other words, 
\begin{align}
\Pi^{ML}_{i,j} = P(\Pi^{AD}_{i,j} | \textbf{p}_{i,j}) \Pi^{AD}_{i,j} + P(\Pi^{SM}_{i,j} | \textbf{p}_{i,j}) \Pi^{SM}_{i,j}
\end{align}
where $\Pi^{AD}_{i,j}$ and $\Pi^{SM}_{i,j}$ are AD and Smagorinsky predictions for the turbulence model at a point. We note that the same learning framework is deployed in these two conceptually different scenarios. 

\begin{figure}
\centering
\includegraphics[width=0.8\textwidth]{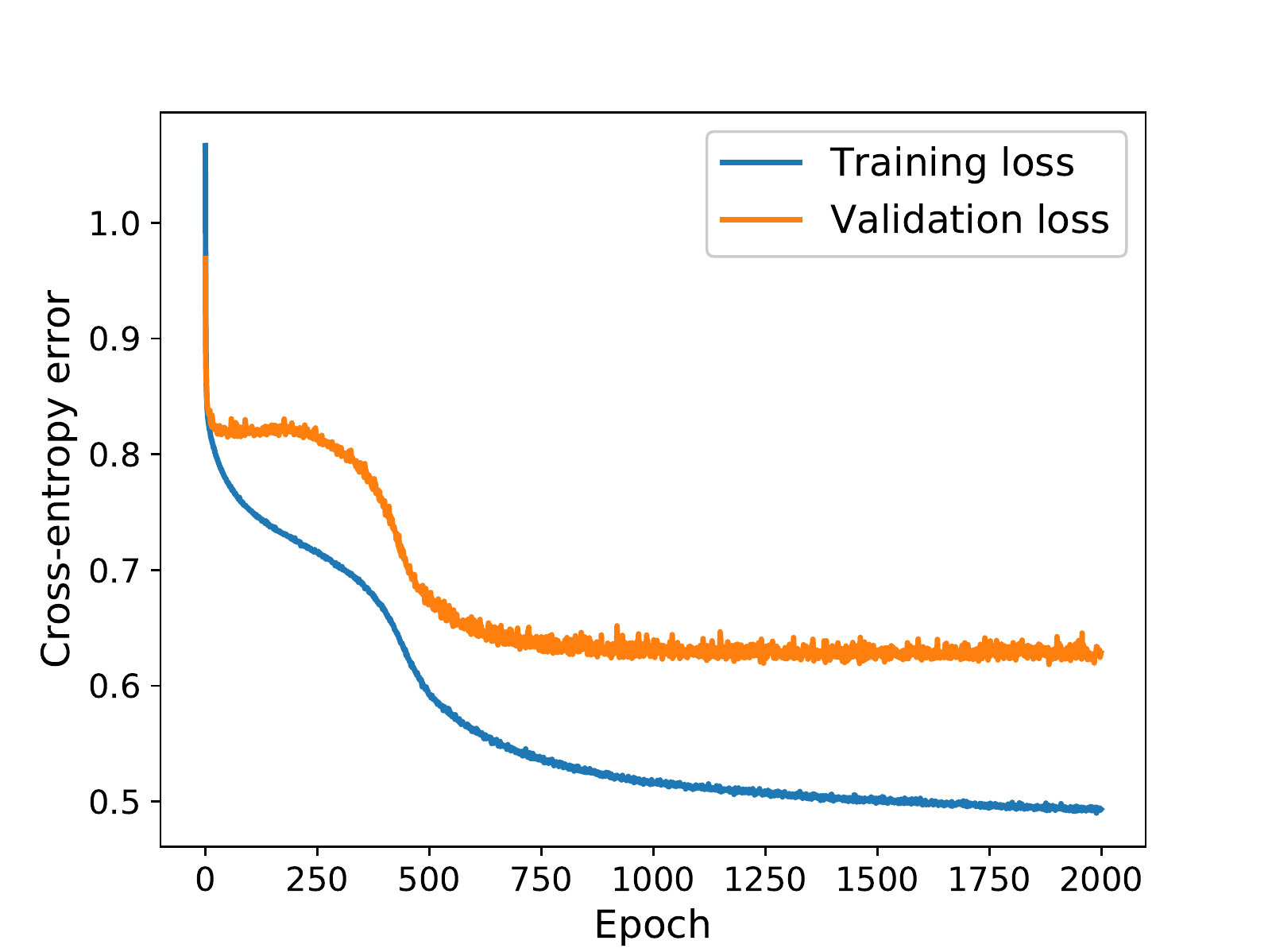}
\caption{Neural-network training and validation loss for the proposed learning framework showing convergence at around 1500 epochs. The best model was chosen according to lowest validation loss for reduced over-fitting in forward deployments.}
\label{Fig3}
\end{figure}

The framework is trained using the previously mentioned categorical cross-entropy error minimization for the one-hot encoded targets. A three-fold cross-validation is utilized with a grid search for the number of layers (between 1 to 8) and number of neurons (between 10 to 100 at intervals of 10) to arrive at the optimal architecture mentioned previously. This optimal network is then trained for 2000 epochs to arrive at a classification accuracy of 79\% for training and approximately 68\% accuracy for validation. Convergence in validation loss was observed at around 1500 epochs as shown in Figure \ref{Fig3}. We note here that our validation data (amounting to one-third of the total training data set) was not exposed to the network during gradient calculation in the back-propagation based training procedure. Effectively, our learning is derived from two-thirds of the total training data while our best-model is chosen from that with the lowest validation loss. This is to ensure that the chances of network extrapolation are minimized. The optimal learning is then deployed into \emph{a posteriori} evolution of the Kraichnan turbulence test case where a pointwise closure deployment is performed for a variety of test cases. We also note that our labeled data is pre-processed to ensure that an equal number of samples are available from each classification regime to prevent our learning from prioritizing one outcome over the other two. 

\section{Results}

We proceed by examining the performance of our framework for various \emph{a posteriori} deployments which act as a rigorous testing of our learning for both classification and blending. We remind the reader that \emph{a posteriori} deployments of learning frameworks imply performance assessments in the presence of challenging numerical errors and represent the ultimate test of a data-driven framework. Briefly, the Kraichnan turbulence problem is specified by periodic boundary conditions on a rectangular domain and an initial condition which is given by an energy spectrum in wave number space. In this two-dimensional problem very fine scales are developed quickly and this leads to the classical $k^{-3}$ scaling of the kinetic energy spectra which is a characteristic of the cascade of enstrophy in two-dimensional turbulence. The turbulence then decays gradually over time and thus represents an unsteady closure modeling assessment for our proposed framework.

We assess the viability of the proposed framework through energy spectra calculations of various reduced-order deployments as well as vorticity structure functions obtained from the same. Time histories of the turbulent kinetic energy (denoted $TKE$) and the variance of vorticity (denoted $\sigma^2 (\bar{\omega})$) are also plotted for forward deployments. Detailed explanations of the numerical schemes and energy spectra calculations utilized for this problem may be found in \cite{maulik2017stable}. Briefly, all our spatial numerical schemes are second-order accurate and our time-integration is third-order total-variation diminishing. Our vorticity structure function calculations are given by \cite{grossmann1992structure}:
\begin{align}
S_\omega = \langle |\bar{\omega}(\textbf{x+r}) - \bar{\omega}(\textbf{x})|^2 \rangle ,
\end{align}
where the angle-brackets indicate ensemble averaging and $\textbf{x}$ indicates a position on the grid with $\textbf{r}$ being a certain distance from this location. Our turbulent kinetic energy is given by
\begin{align}
TKE = \mu (u_f^2 + v_f^2),
\end{align}
where $u_f$ and $v_f$ are fluctuating quantities given by 
\begin{align}
u_f = \bar{u} - \mu (\bar{u}) \\
v_f = \bar{v} - \mu (\bar{v}),
\end{align}
and where $\mu(a)$ implies the spatial mean of a field variable $a$. We note that the components of velocity $u,v$ are computed by second-order accurate central finite-difference implementations of 
\begin{align}
\bar{u} &= \frac{\partial \bar{\psi}}{\partial y}, \quad \bar{v} = -\frac{\partial \bar{\psi}}{\partial x}.
\end{align}
In a similar manner the variance of vorticity, at each time step, is computed using 
\begin{align}
\sigma^2 (\bar{\omega}) = \mu \left((\bar{\omega}-\mu(\bar{\omega}))^2 \right).
\end{align}

In all the following assessments, the proposed framework is denoted as ML (and specified to be deployed as a classifier or a blender) and it is compared to the AD and DS models. We remind the reader that the framework utilizes the \emph{static} Smagorinsky model (denoted SM) with $C_s=1.0$ within its formulation but is assessed against the DS approach. The reader may note that the value of $C_s=1.0$ proves over-dissipative for this particular test case as shown in \cite{maulik2019subgrid}.

\subsection{Model classification}
\label{model_class}

\begin{figure}
\centering
\mbox{
\subfigure{\includegraphics[width=0.48\textwidth]{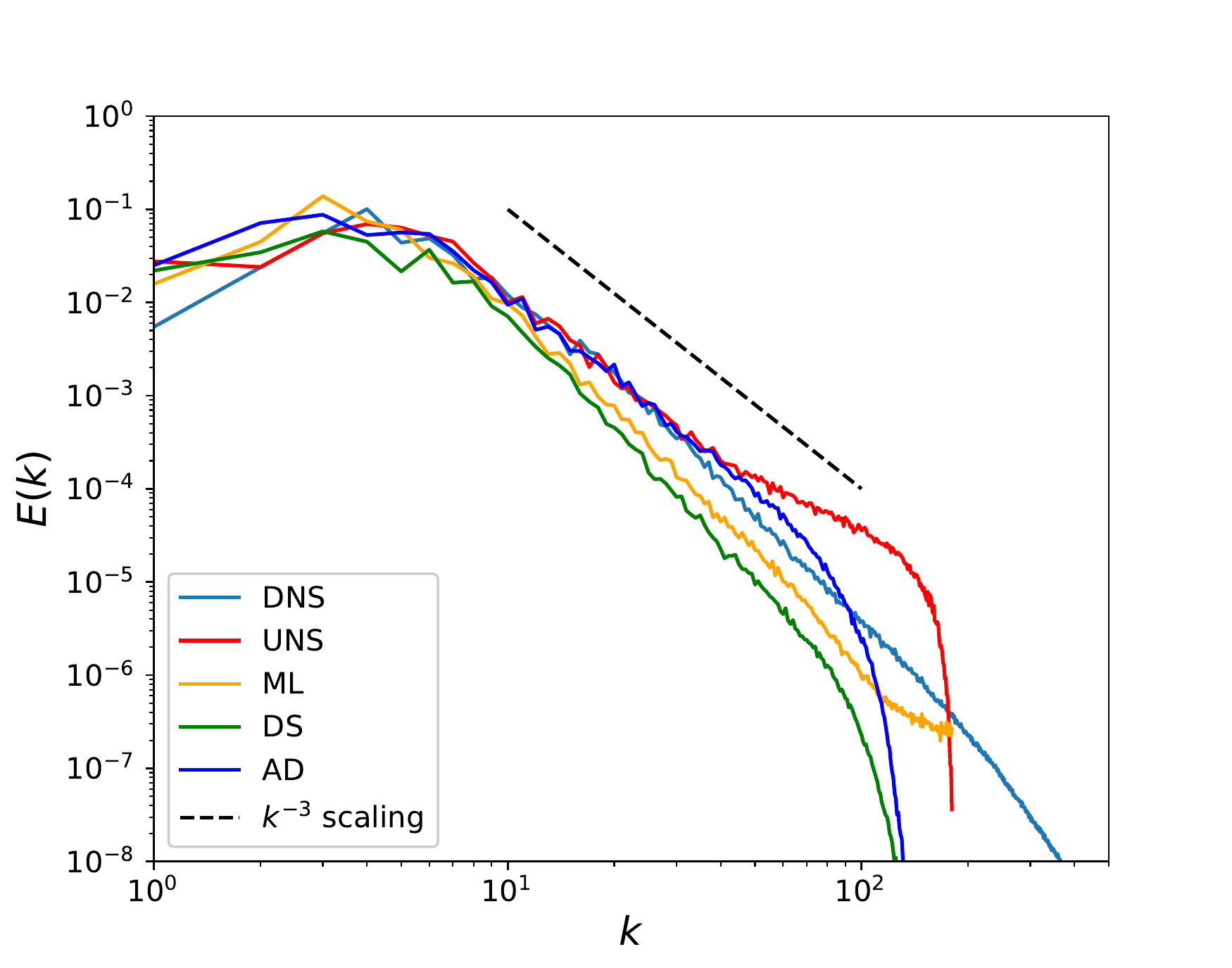}}
\subfigure{\includegraphics[width=0.48\textwidth]{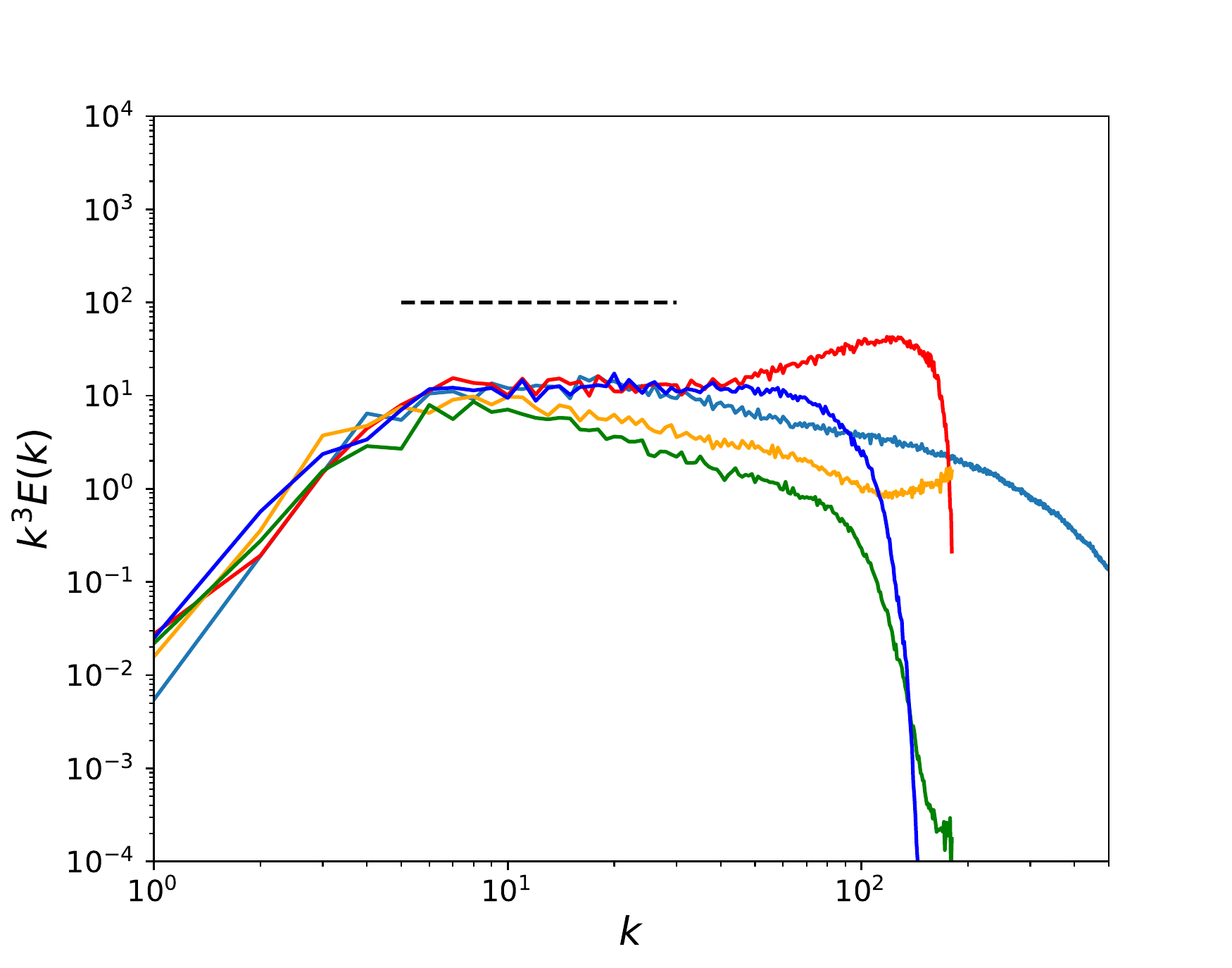}}
}
\caption{\emph{A posteriori} kinetic-energy spectra (left) and compensated kinetic-energy spectra (right) for $Re=32000$ at $t=4$ and at $N^2=256^2$ degrees of freedom. The proposed framework (deployed as a classifier) balances the dissipative natures of the AD and the DS models.}
\label{Fig4}
\end{figure}

In this section, we deploy our learning framework as a classifier which spatio-temporally switches between three closure modeling hypotheses during flow-evolution. Figure \ref{Fig4} shows the performance of our proposed framework for the forward deployment of the Kraichnan turbulence problem in the form of energy spectra predictions at $t=4$. For comparison, no-model results (denoted UNS), the DS method and AD are also shown along with DNS spectra. One can observe that the classifier balances the dissipative natures of the SM and AD hypothesis to obtain a performance similar to the that of the DS approach. While at the lower wavenumbers, the AD procedure seems to be more accurate in statistical capture, the higher wavenumbers are stabilized adequately by the classifier. We would like to note that that SM hypothesis with $C_s=1.0$ is highly dissipative and this results shows that the classifier avoids its deployment to a large degree for improved \emph{a posteriori} performance. We clarify that for spectral cut-off filtering, FDNS spectra and DNS spectra are identical till the grid cut-off wave number \citep{maulik2018deconvolution}. 

\begin{figure}
\centering
\mbox{
\subfigure{\includegraphics[width=0.48\textwidth]{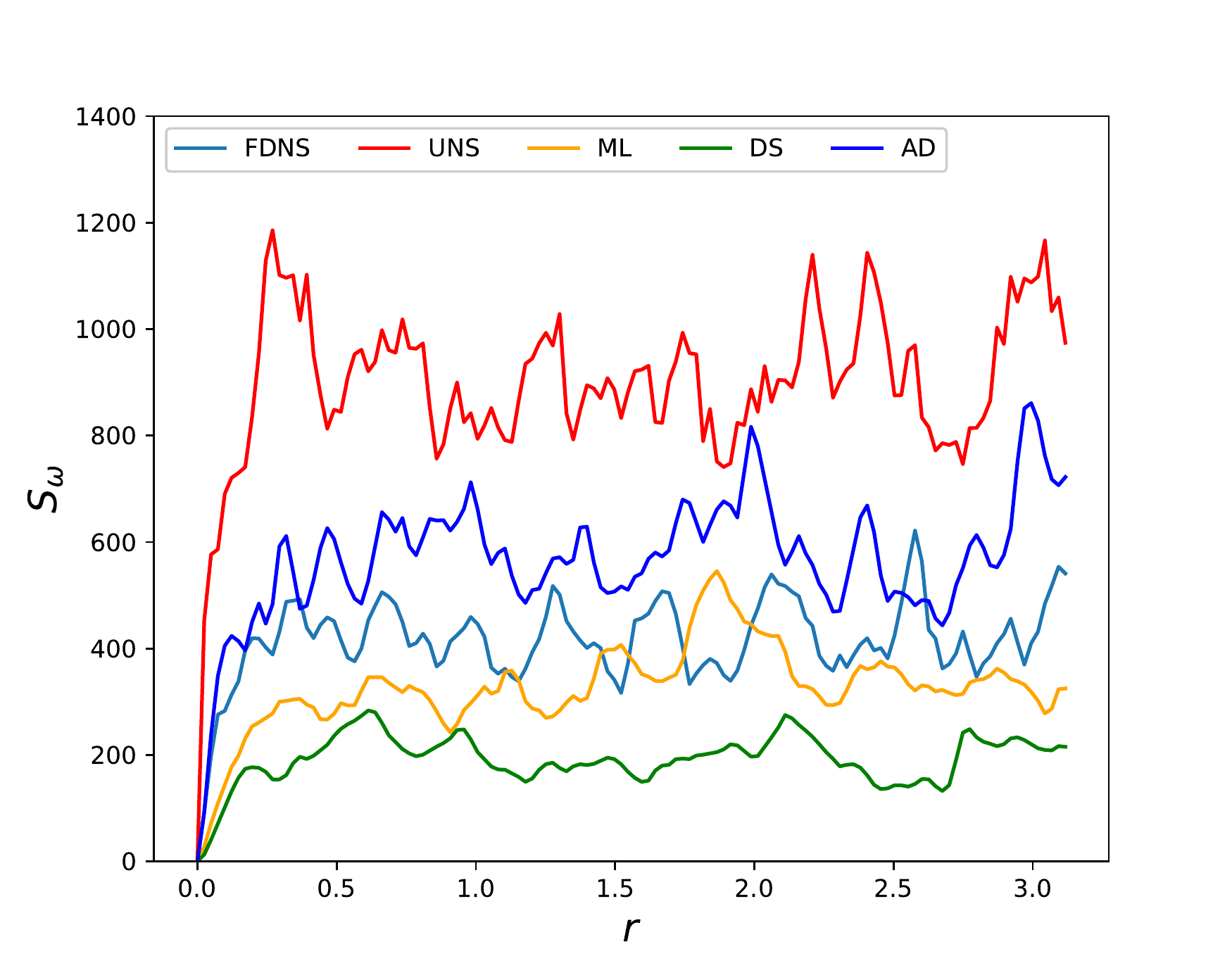}}
\subfigure{\includegraphics[width=0.48\textwidth]{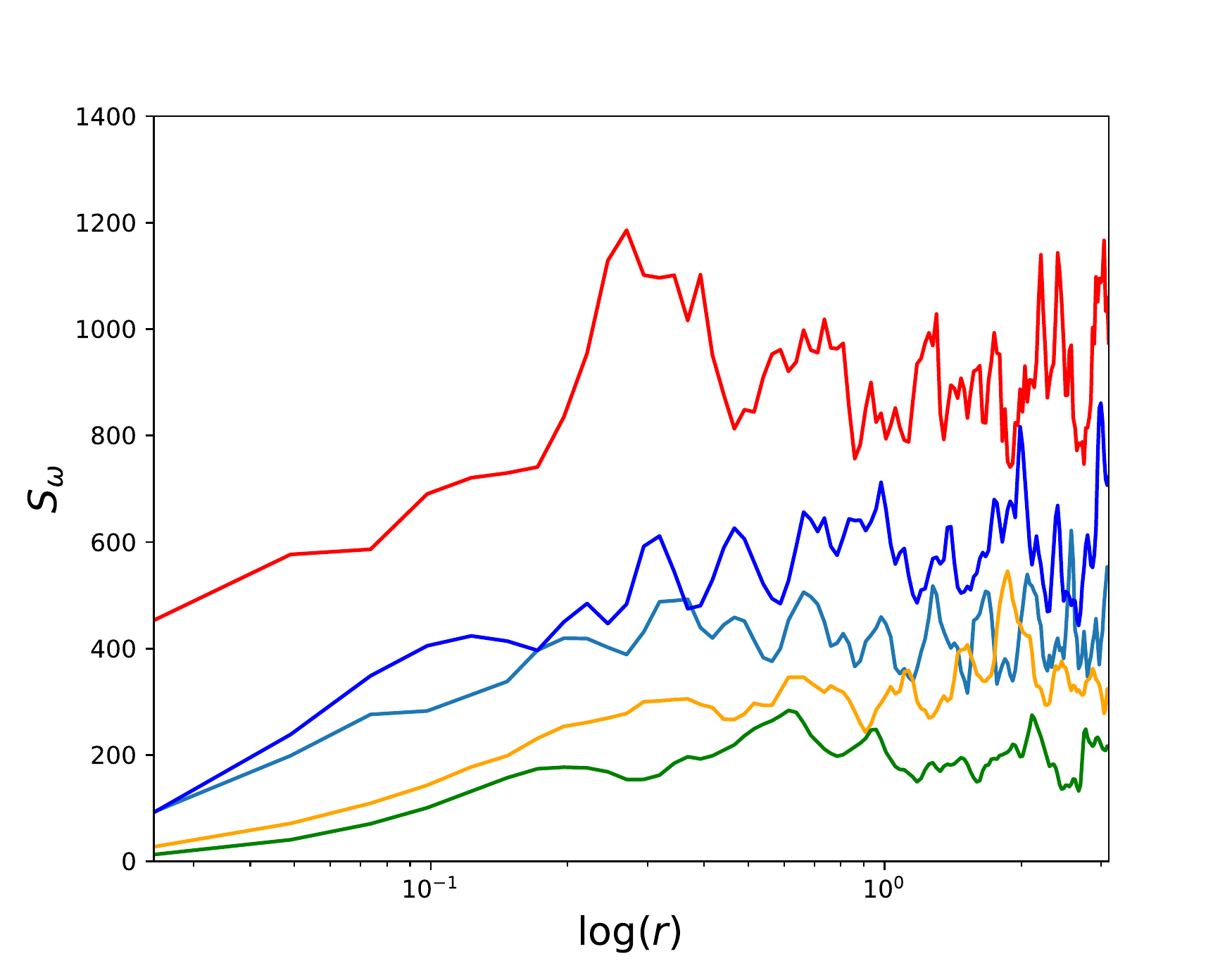}}
}
\caption{\emph{A posteriori} vorticity structure functions plotted against $\textbf{r}$ (left) and $\log(\textbf{r})$ (right) for $Re=32000$ at $t=4$ and at $N^2=256^2$ degrees of freedom. It is observed that AD performs better in the near-region whereas the proposed framework behaves similar to the DS approach.}
\label{Fig5}
\end{figure}

\begin{figure}
\centering
\mbox{
\subfigure{\includegraphics[width=0.48\textwidth]{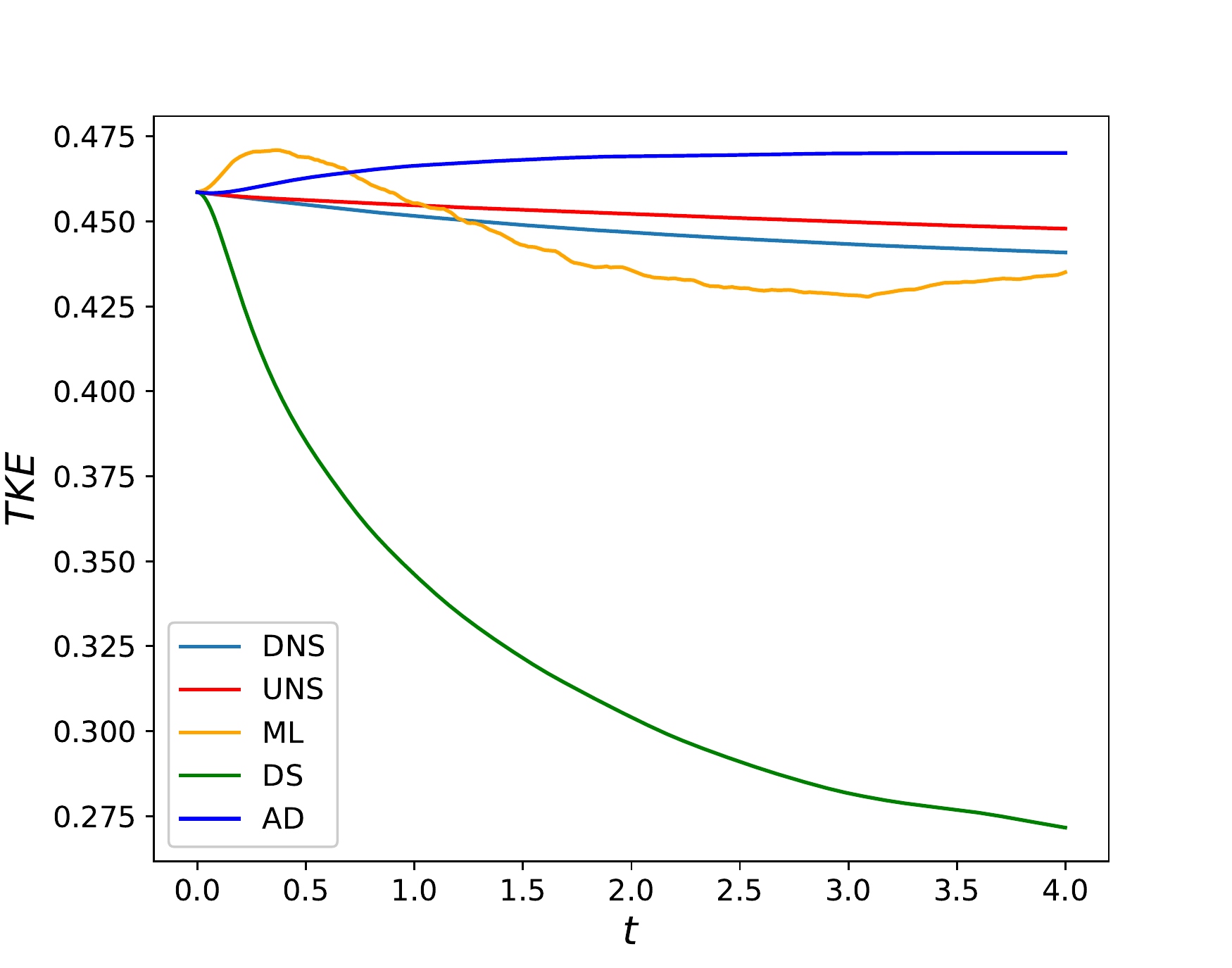}}
\subfigure{\includegraphics[width=0.48\textwidth]{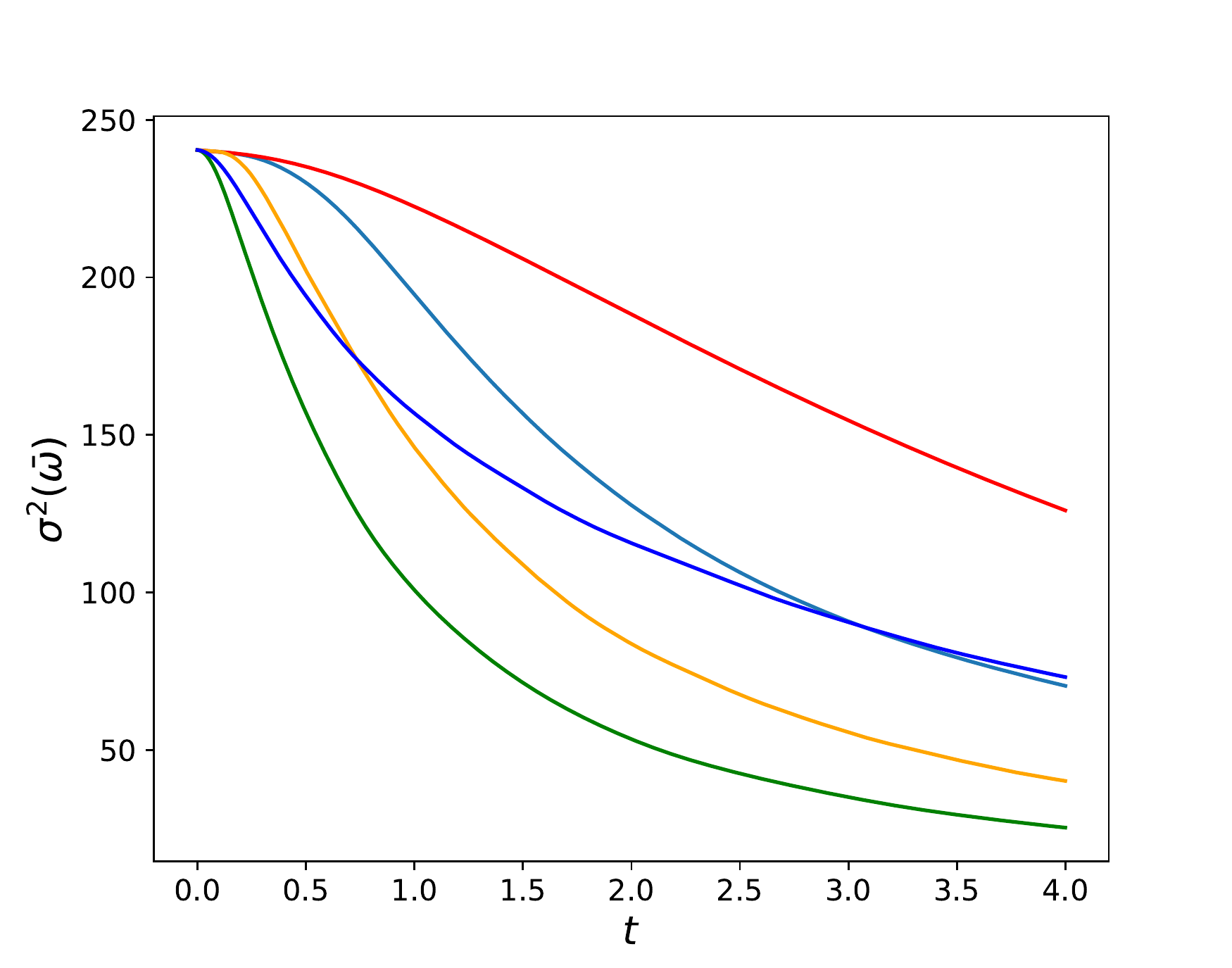}}
}
\caption{Time-histories for turbulent kinetic energy (left) and vorticity variance (right) for $Re=32000$ at $N^2=256^2$ degrees of freedom. The proposed method can be seen to adapt between the behavior of the AD and DS techniques.}
\label{Fig6}
\end{figure}

Figure \ref{Fig5} details vorticity structure function assessments in our domain where assessments with FDNS show that the proposed framework is adequately capable of stabilizing turbulence correlations at $t=4$. We note that the structure functions are predicted more accurately by AD at low values of $\textbf{r}$ whereas the proposed classifier behaves similar to a DS implementation thereby indicating a dynamic dissipation on the grid. It may be so that the adaptive dissipation prioritizes noise removal and thus introduces errors at low values of $\textbf{r}$ as seen through stable structure functions at saturation (i.e., at higher values of $\textbf{r}$). A further assessment is deployed in the form of time-histories of TKE and $\sigma^2 (\bar{\omega})$ as shown in Figure \ref{Fig6}. Once again, the classifier is seen to have a varying trend in TKE predictions compared to the AD and DS techniques indicating varying dissipation strengths. The vorticity variance predictions are also seen to be balanced between that of the DS and AD models indicating the balance of dissipative tendencies. 

\begin{figure}
\centering
\mbox{
\subfigure{\includegraphics[width=0.48\textwidth]{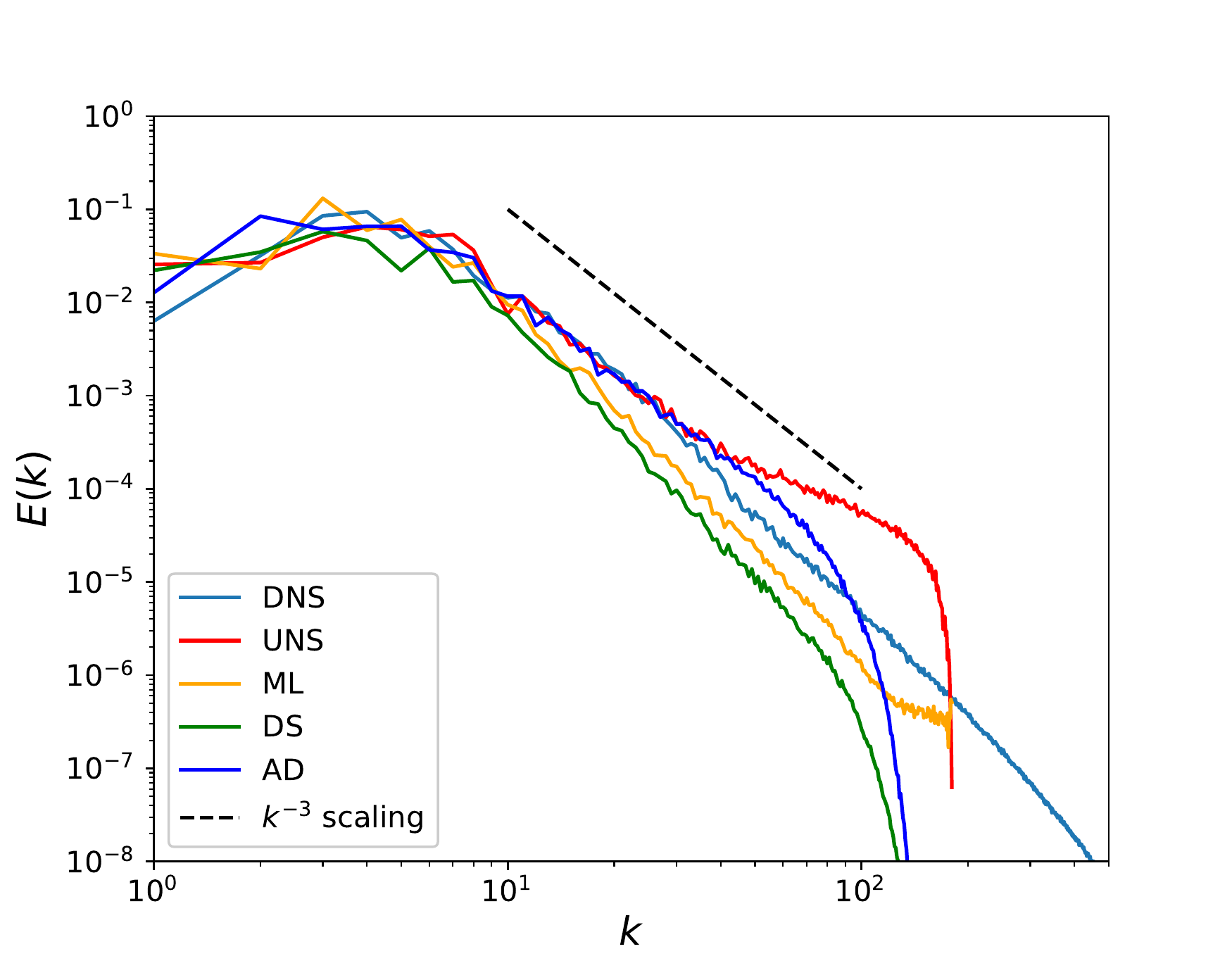}}
\subfigure{\includegraphics[width=0.48\textwidth]{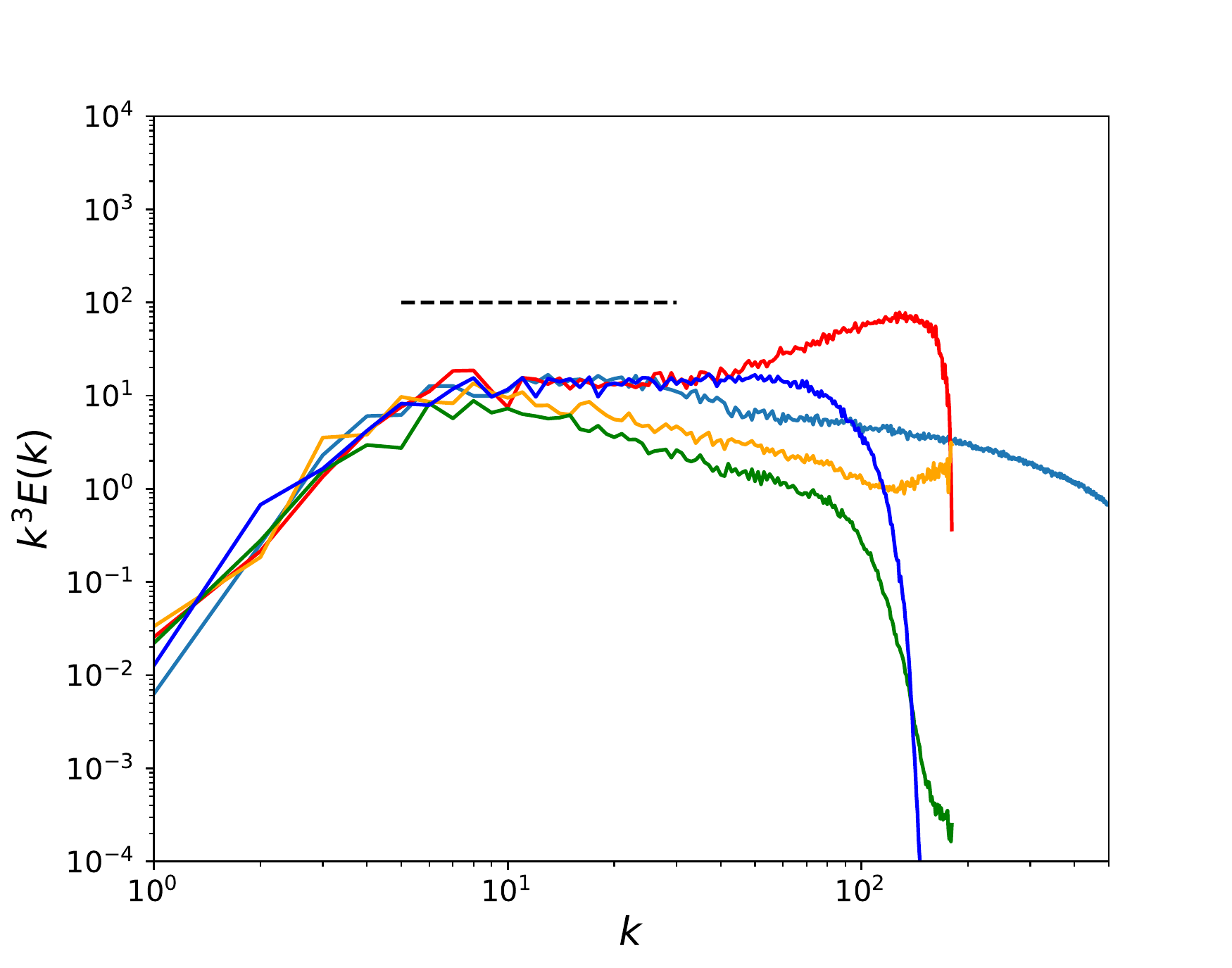}}
}
\caption{\emph{A posteriori} kinetic-energy spectra (left) and compensated kinetic-energy spectra (right) for $Re=64000$ at $t=4$ and at $N^2=256^2$ degrees of freedom. This assessment displays closure effectiveness for a Reynolds number not utilized in the training data.}
\label{Fig7}
\end{figure}

\begin{figure}
\centering
\mbox{
\subfigure{\includegraphics[width=0.48\textwidth]{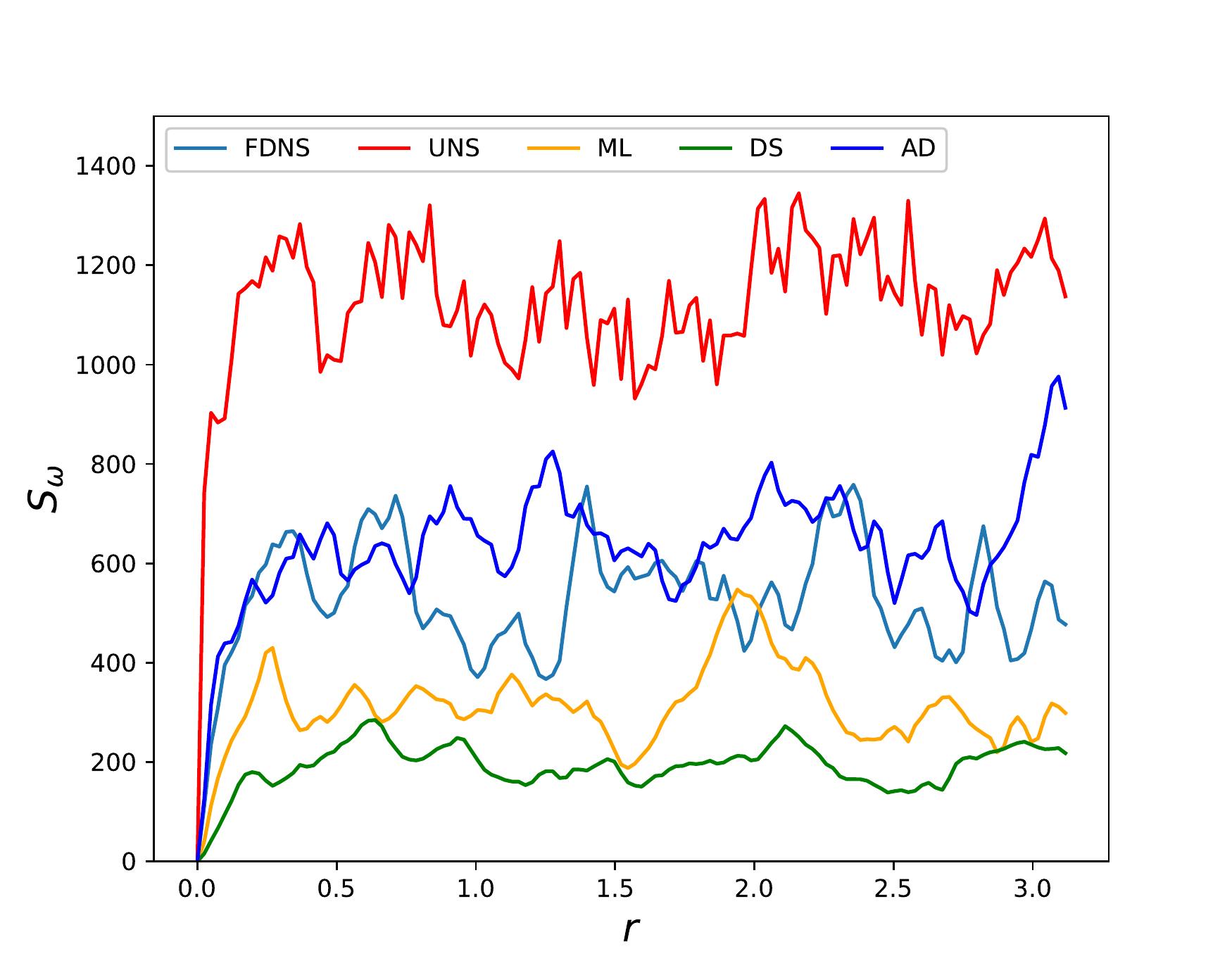}}
\subfigure{\includegraphics[width=0.48\textwidth]{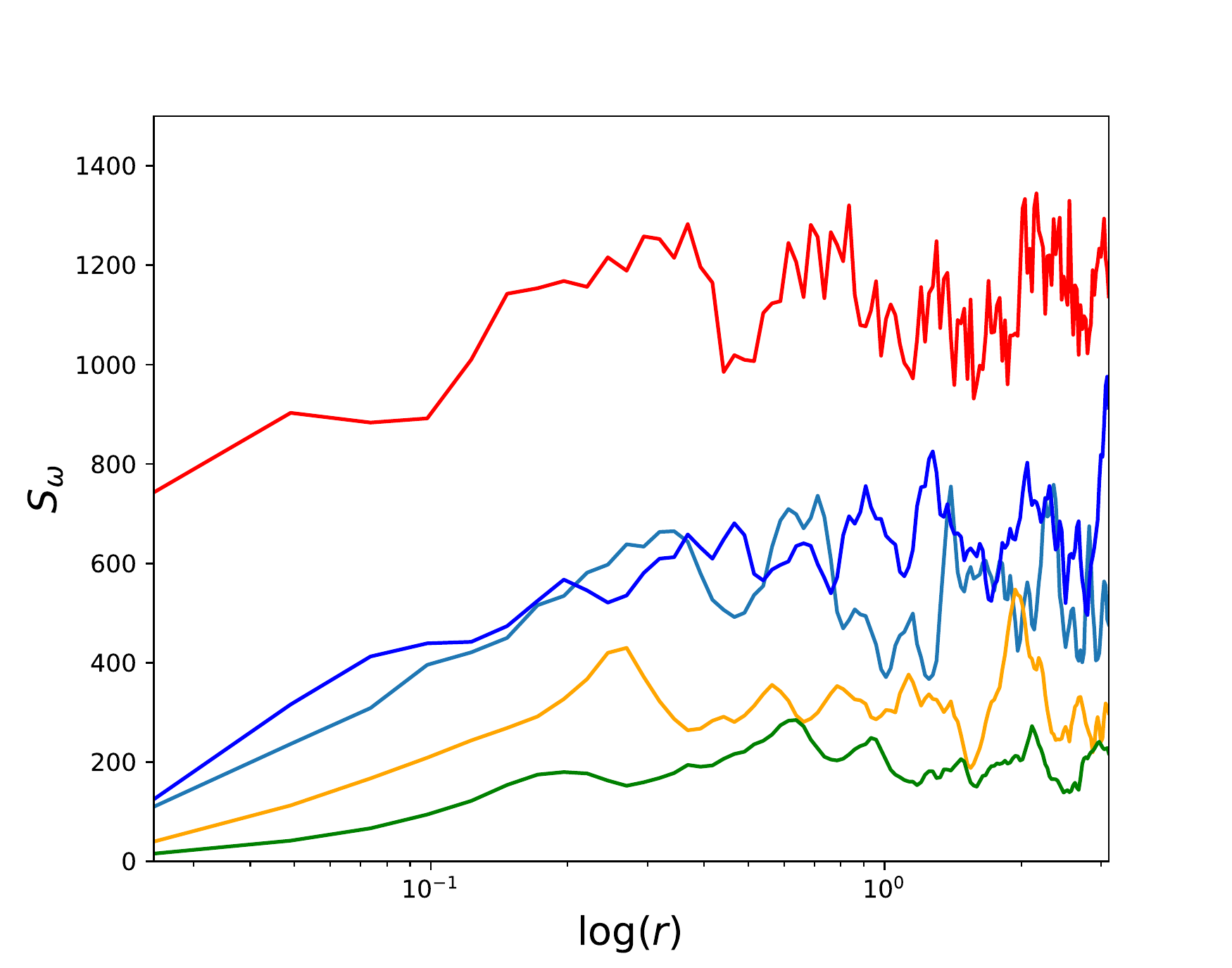}}
}
\caption{\emph{A posteriori} vorticity structure functions plotted against $\textbf{r}$ (left) and $\log(\textbf{r})$ (right) for $Re=64000$ at $t=4$ and at $N^2=256^2$ degrees of freedom. It is observed that solely AD performs better in the near-region whereas the proposed framework behaves similar to the DS approach. The behavior is similar to that observed for within training data regime deployment.}
\label{Fig8}
\end{figure}

\begin{figure}
\centering
\mbox{
\subfigure{\includegraphics[width=0.48\textwidth]{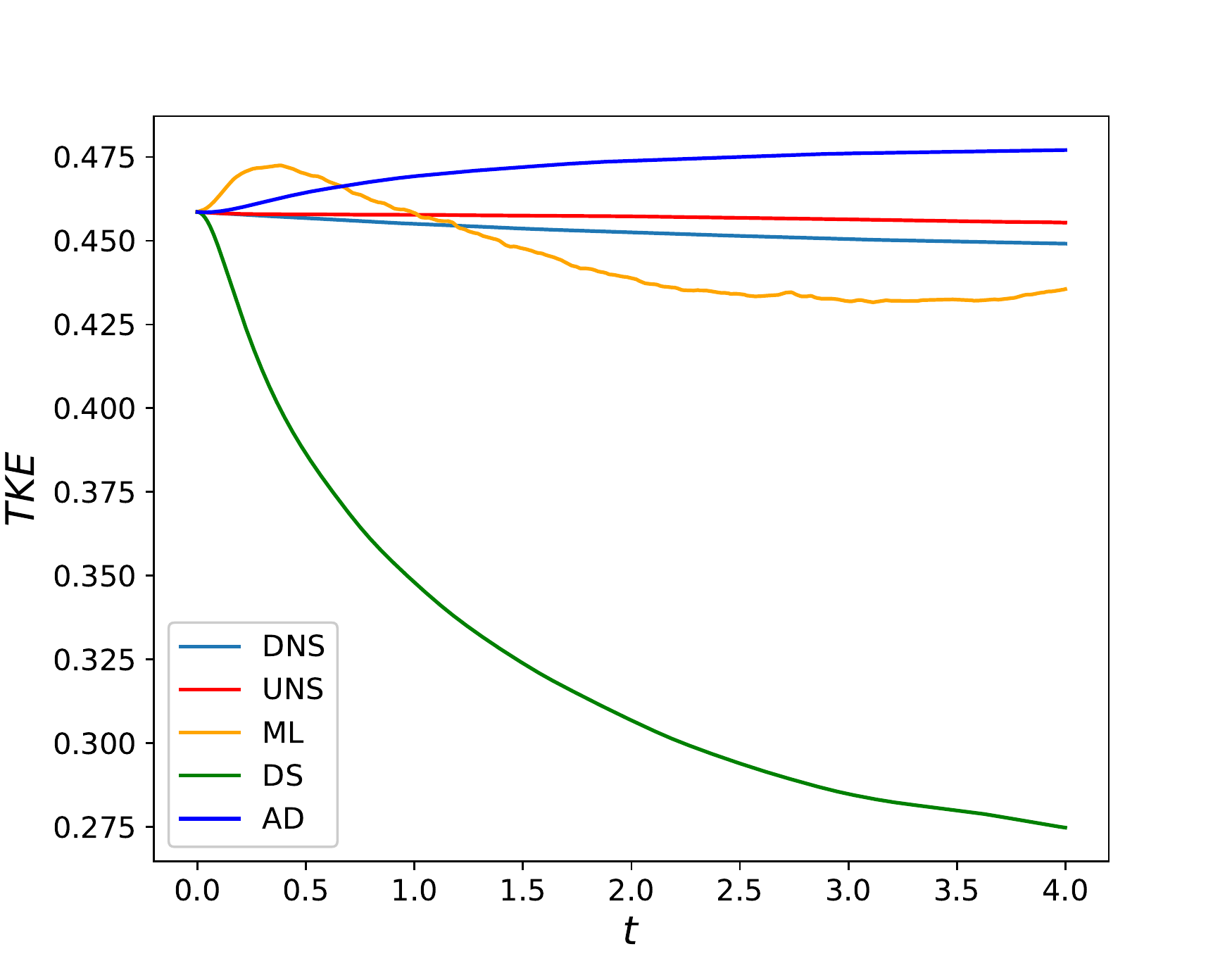}}
\subfigure{\includegraphics[width=0.48\textwidth]{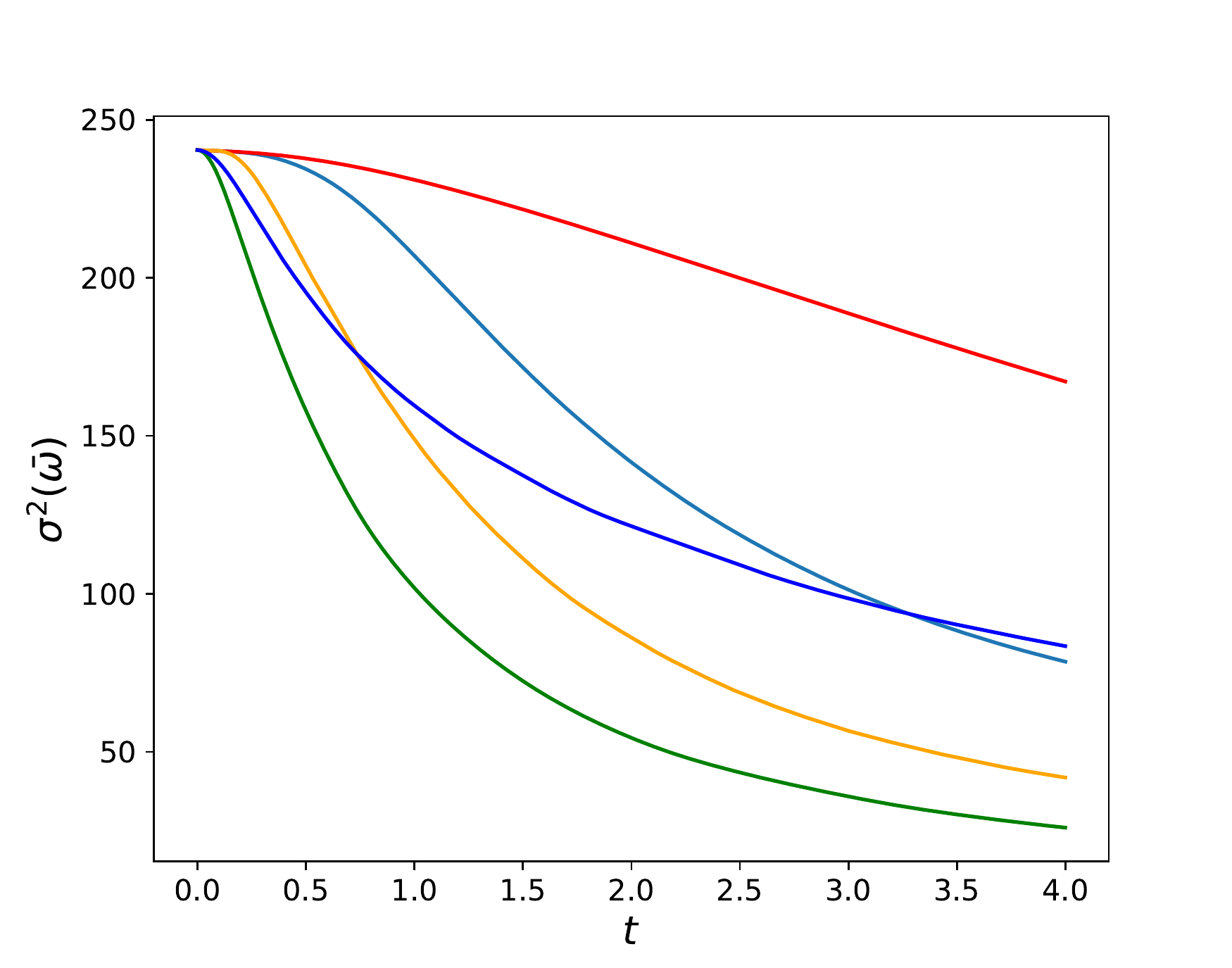}}
}
\caption{Time-histories for turbulent kinetic energy (left) and vorticity variance (right) for $Re=64000$ at $N^2=256^2$ degrees of freedom. The proposed method can be seen to adapt between the behavior of the AD and DS techniques and acts as an additional validation for deployment to different Reynolds numbers.}
\label{Fig9}
\end{figure}

We proceed by performing a thorough validation of our learning framework by assessing its performance for prediction tasks that it has not been exposed to in training. This is established by testing closure efficiency for a Reynolds number of 64000. We remind the reader that map optimization was performed solely for $Re=32000$ and this represents an additional validation of the learning. Kinetic energy spectra for this experiment are shown in Figure \ref{Fig7} where it is observed that the classifier performs in a very similar fashion to the $Re=32000$ test-case with AD performing more efficiently at the lower wavenumbers of the inertial range but the ML approach stabilizing high-wavenumber noise effectively. This indicates that the learning has generalized, atleast on the current degree of coarse-graining. We also perform additional assessments such as those shown in Figure \ref{Fig8} and Figure \ref{Fig9}. The former shows the vorticity structure function trends for this out-of-training range learning assessment and the latter shows the time-histories of TKE and $\sigma^2(\bar{\omega})$. Very similar trends for both these assessments are obtained when compared to the $Re=32000$ test-case with time varying trends in TKE and vorticity variance capture. 

\begin{figure}
\centering
\mbox{
\subfigure{\includegraphics[width=0.48\textwidth]{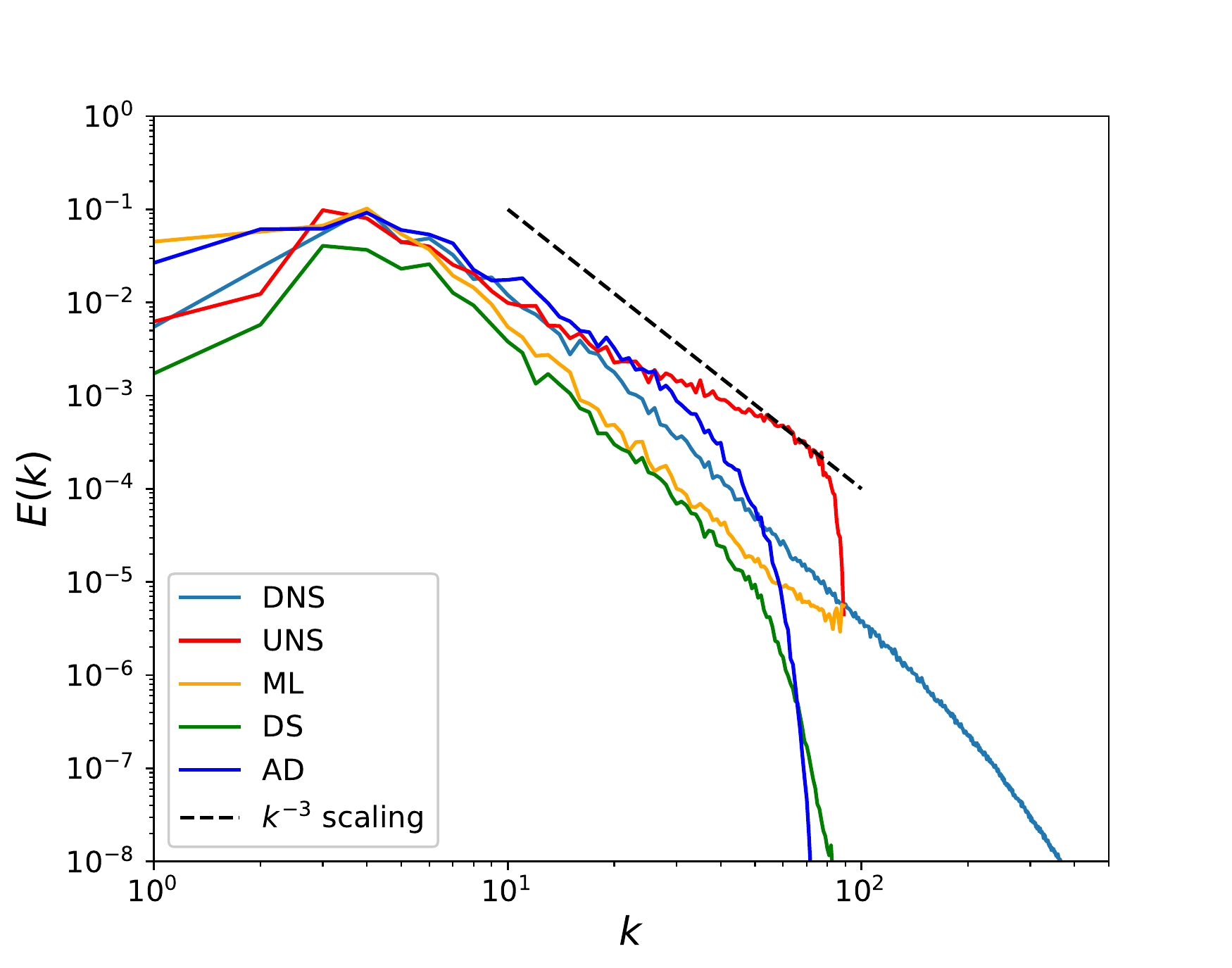}}
\subfigure{\includegraphics[width=0.48\textwidth]{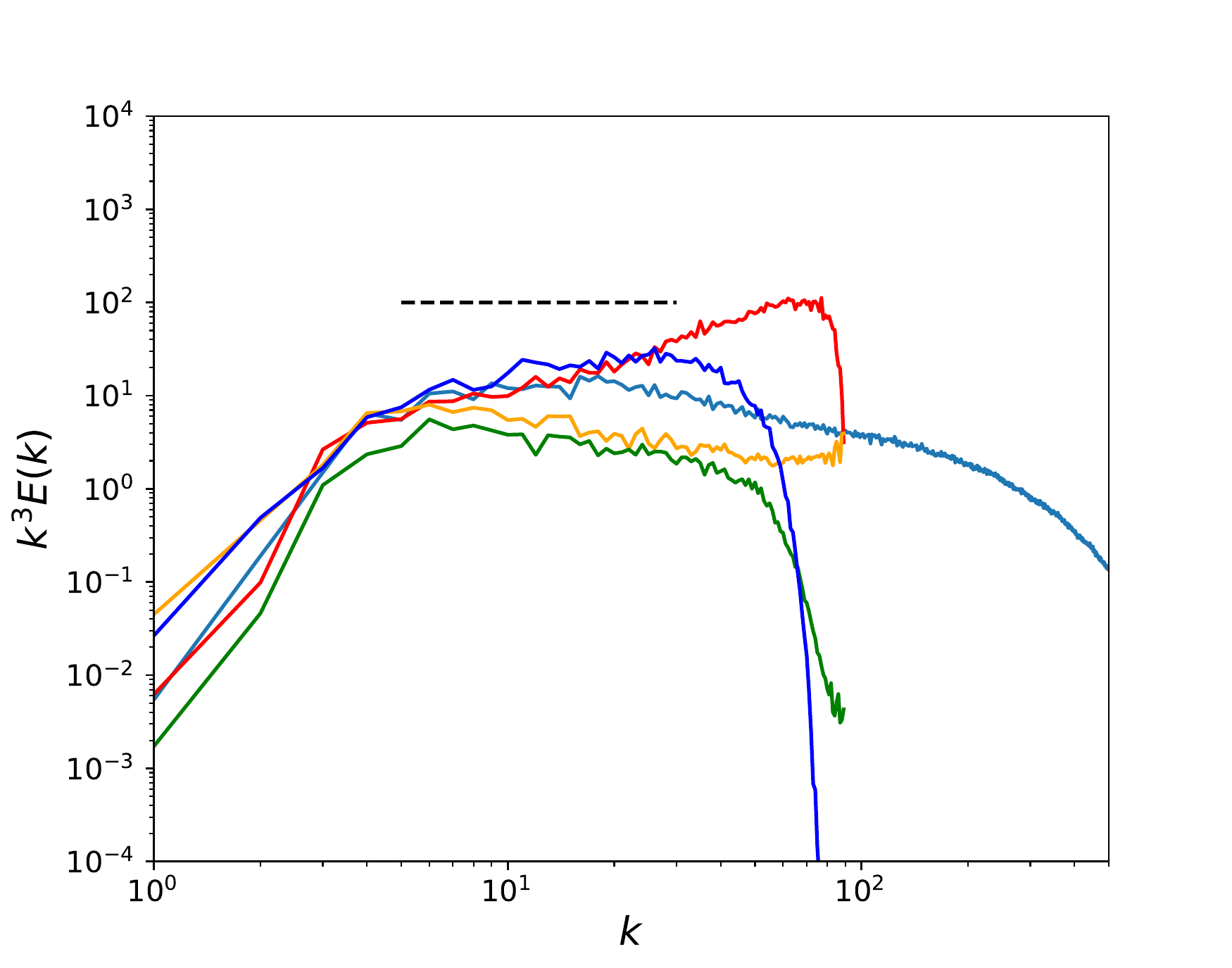}}
}
\caption{\emph{A posteriori} kinetic-energy spectra (left) and compensated kinetic-energy spectra (right) for $Re=32000$ at $t=4$ and at $N^2=128^2$ degrees of freedom. This assessment displays closure effectiveness for a coarse-grained resolution not utilized in the training data.}
\label{Fig10}
\end{figure}

\begin{figure}
\centering
\mbox{
\subfigure{\includegraphics[width=0.48\textwidth]{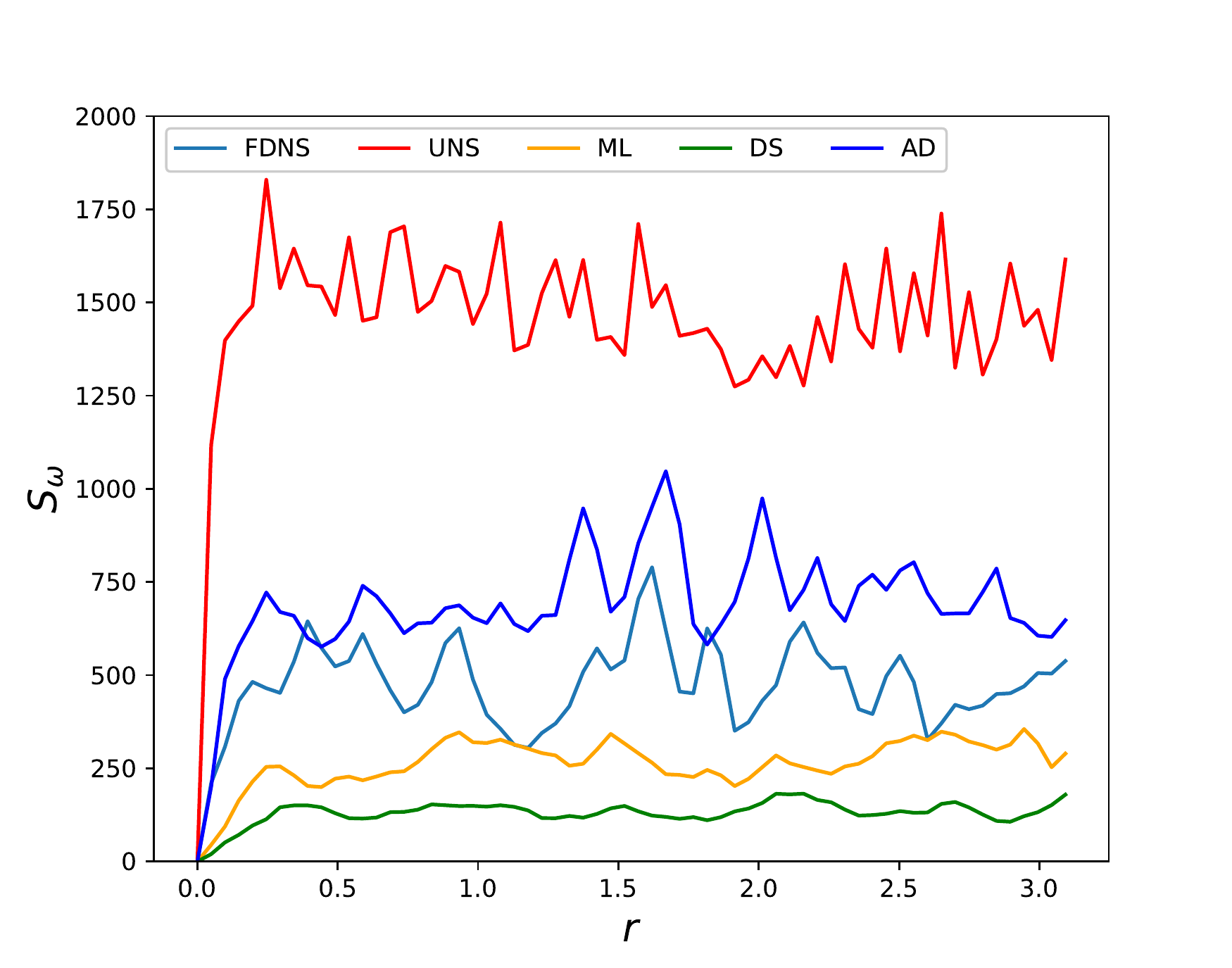}}
\subfigure{\includegraphics[width=0.48\textwidth]{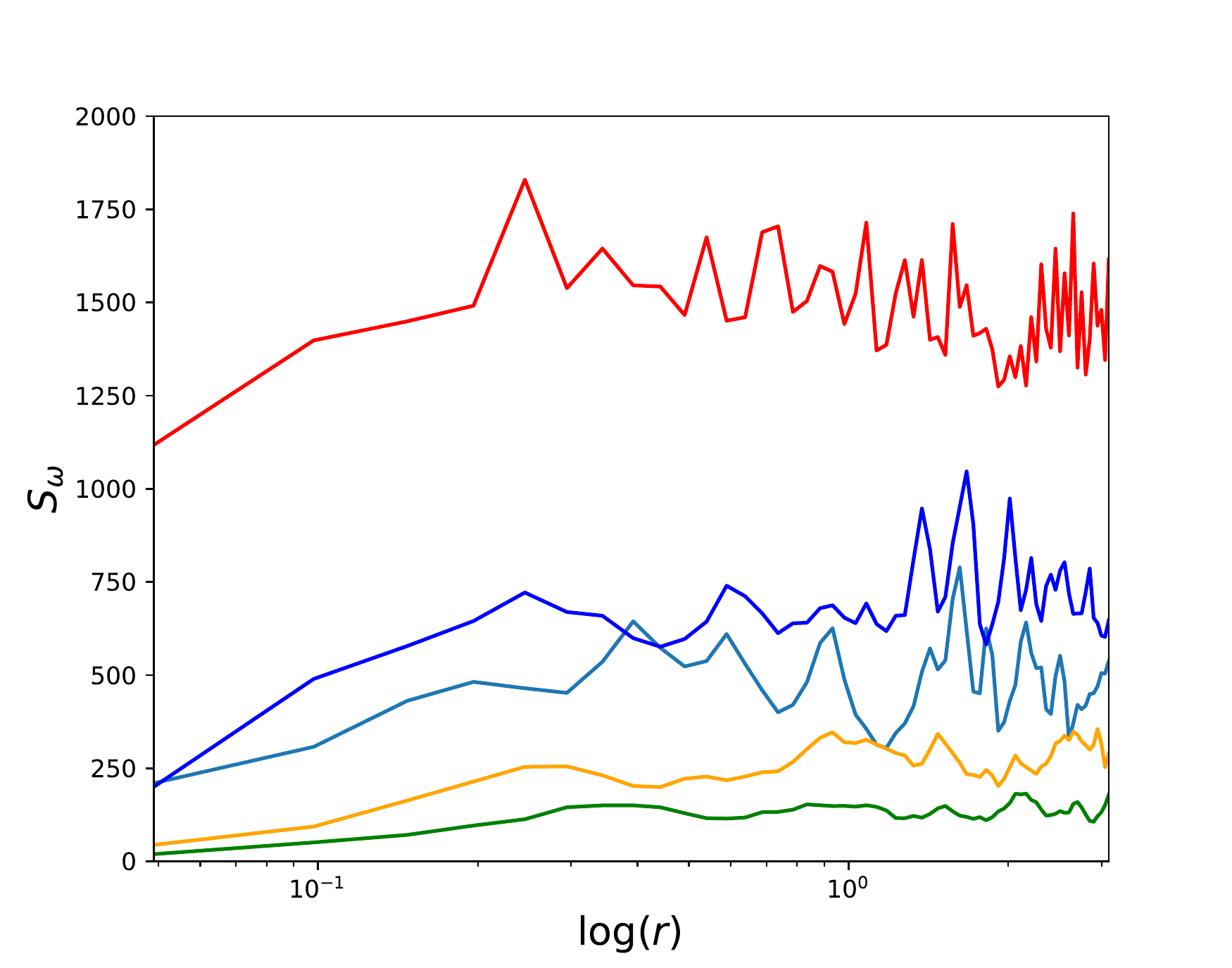}}
}
\caption{\emph{A posteriori} vorticity structure functions plotted against $\textbf{r}$ (left) and $\log(\textbf{r})$ (right) for $Re=32000$ at $t=4$ and at $N^2=128^2$ degrees of freedom. It is observed that solely AD performs better in the near-region whereas the proposed framework behaves similar to the DS approach. The behavior is similar to that observed for within training resolution deployment.}
\label{Fig11}
\end{figure}

\begin{figure}
\centering
\mbox{
\subfigure{\includegraphics[width=0.48\textwidth]{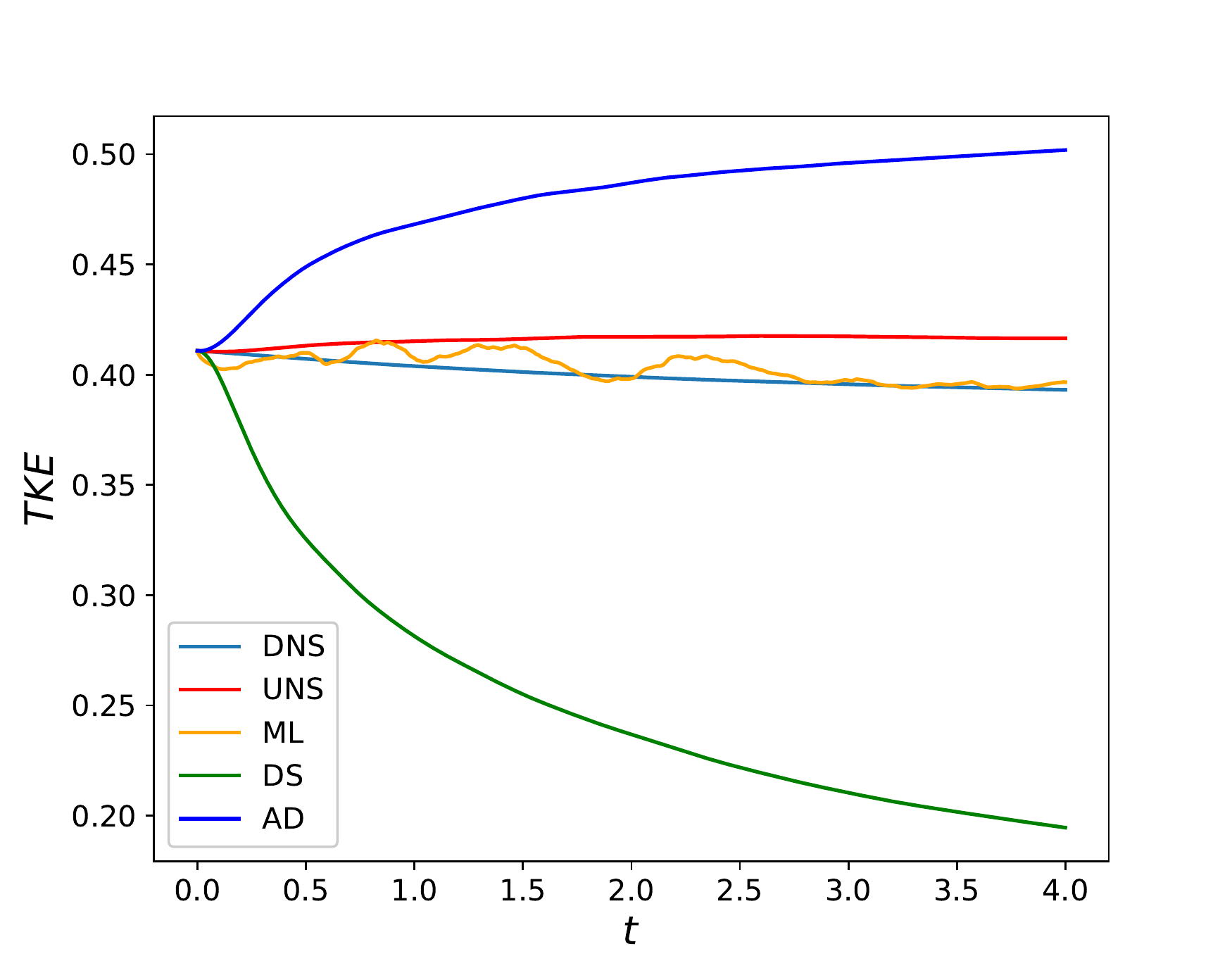}}
\subfigure{\includegraphics[width=0.48\textwidth]{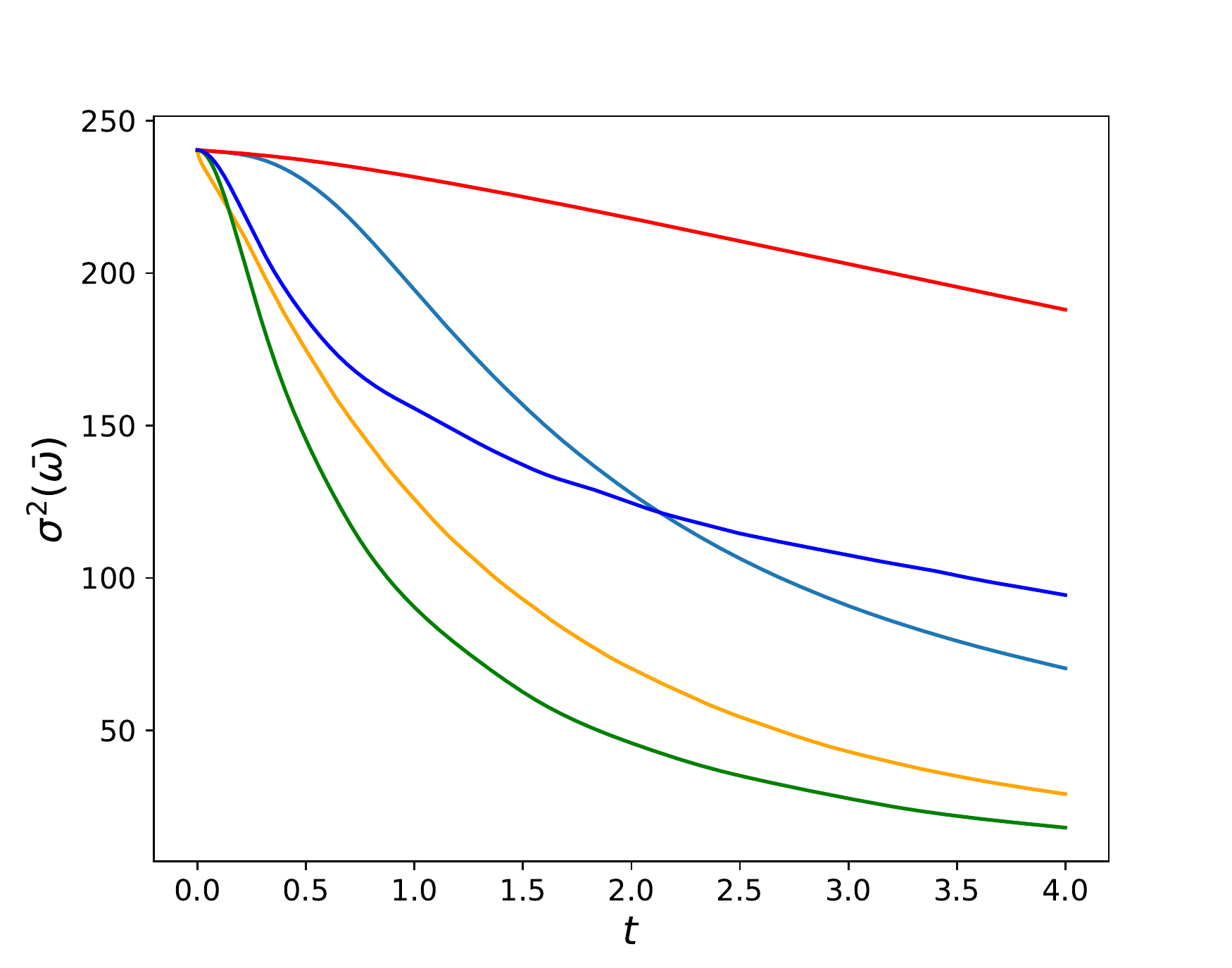}}
}
\caption{Time-histories for turbulent kinetic energy (left) and vorticity variance (right) for $Re=32000$ at $N^2=128^2$ degrees of freedom. The proposed method can be seen to adapt between the behavior of the AD and DS techniques and acts as an additional validation for deployment to similar coarse-grained resolutions.}
\label{Fig12}
\end{figure}

While the aforementioned test-cases validated the learning of the classifier on different control parameters (and flow evolutions) given by the Reynolds number. We proceed by assessing the performance and stability of the classifier on a reduced degree-of-freedom evolution given by $N^2=128^2$. This test was to examine if the classifier could retain a viable learning for deployment on slightly different grid support. Figure \ref{Fig10} shows the kinetic energy spectra for a deployment at this reduced degree-of-freedom at a Reynolds number of 32000. It is observed that the proposed classifier is able to avoid inaccuracies related to AD's lack of dissipation. Indeed, it is well-known that AD requires a sufficiently fine resolution in comparison to the eddy-viscosity hypothesis based models for appropriate utilization of their inverse-filtering \citep{germano2015similarity,guermond2004mathematical}. A similar trend may also be observed in Figure \ref{Fig11} with the vorticity structure functions where once again the AD technique proves accurate at lower distances in comparison the the DS and the ML methods. The ML classifier however is slightly more accurate than the DS approach. The time-histories for TKE and vorticity variance, shown in Figure \ref{Fig12}, display a greater amount of variation in the classification framework with TKE values oscillating but remaining close to the DNS results. It must be noted that the no-model and AD hypothesis prevent the classifier from going into a fully SM deployment which is highly dissipative. This explains the similarity with DS results in terms of spectra and vorticity-variance. 

In addition to the test-case with a slightly reduced grid-resolution in comparison to training data generation. We also perform a grid-dependence check on the accuracy of our classification framework as shown in Table \ref{Table1}. We perform a hypothesis segregation (as introduced previously) to label all points on a coarse-grid with an optimal closure hypothesis and assess if the learning at $N^2=256^2$ is able to categorize them appropriately. It can be seen that accuracies around the same resolution as that of the training data are approximately similar to validation accuracy during network optimization. However, on intense coarse-graining, accuracies are seen to drop significantly. However, we note that even at the coarsest resolution of $N^2=32^2$, accuracies greater than 33\% indicate some form of learning retention. 

\begin{table}
\begin{center}
\begin{tabular}{cccccc}
Time    & $N^2=512^2$ & $N^2=256^2$ & $N^2=128^2$ & $N^2=64^2$ & $N^2=32^2$  \\ \hline
$t = 1$ & 65.77 & 63.04 & 56.51 & 52.17 & 47.65 \\
$t = 2$ & 60.89 & 60.47 & 61.02 & 55.62 & 41.99 \\
$t = 3$ & 68.05 & 65.08 & 61.54 & 53.32 & 46.29 \\
$t = 4$ & 63.93 & 66.04 & 60.24 & 48.33 & 48.54
\end{tabular}
\caption{Classification accuracy percentages for different grid-resolutions in \emph{a priori} to illustrate how accurately our base learning can predict correct labels. Accuracies are seen to drop when resolutions are coarsened radically. However, some learning is retained as evidenced by accuracies greater than 33\%.}
\label{Table1}
\end{center}
\end{table}

\begin{figure}
\centering
\mbox{
	\subfigure[$Re=32000$, $N^2=256^2$]{\includegraphics[width=0.48\textwidth]{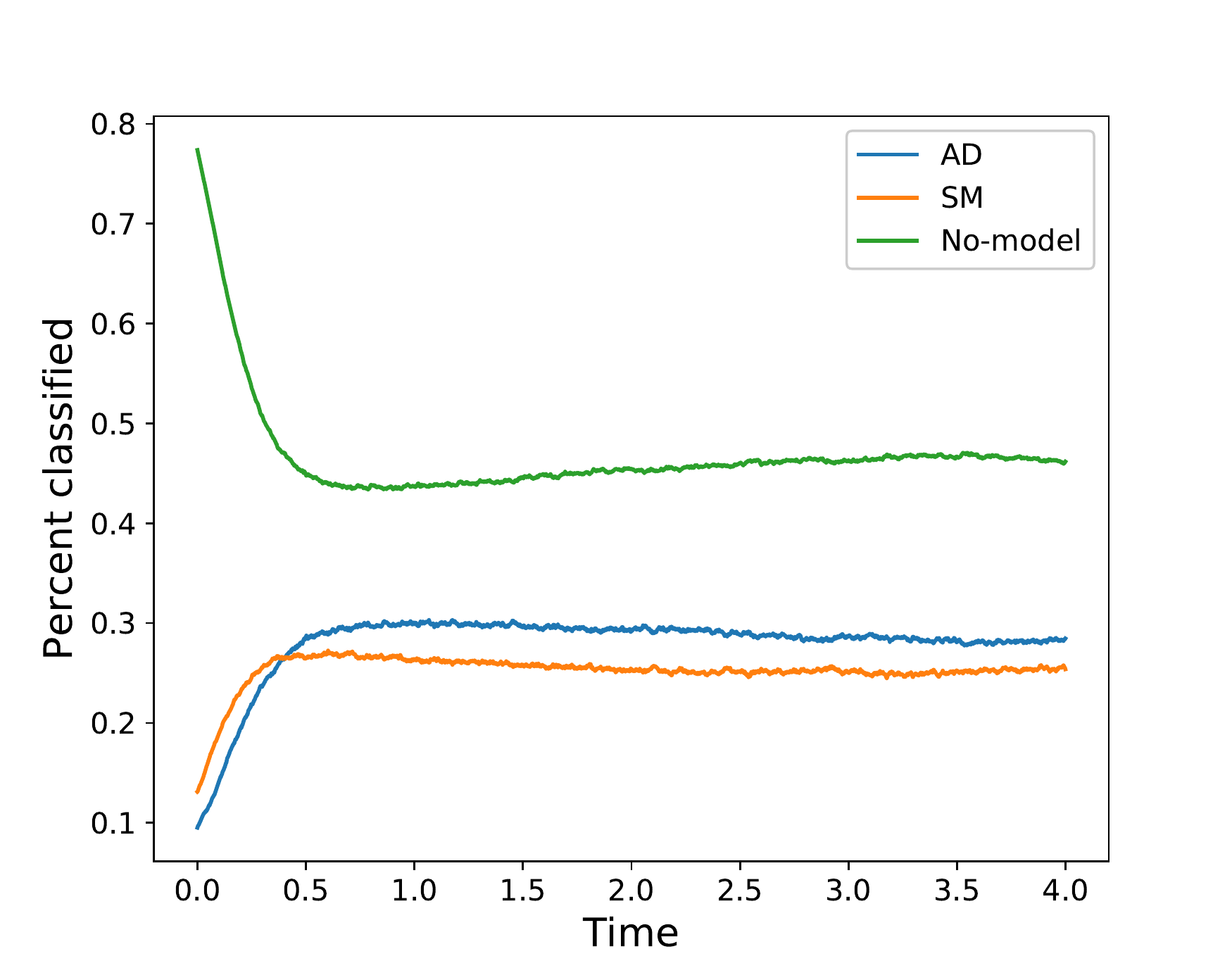}}
	} \\
\mbox{
	\subfigure[$Re=64000$, $N^2=256^2$]{\includegraphics[width=0.48\textwidth]{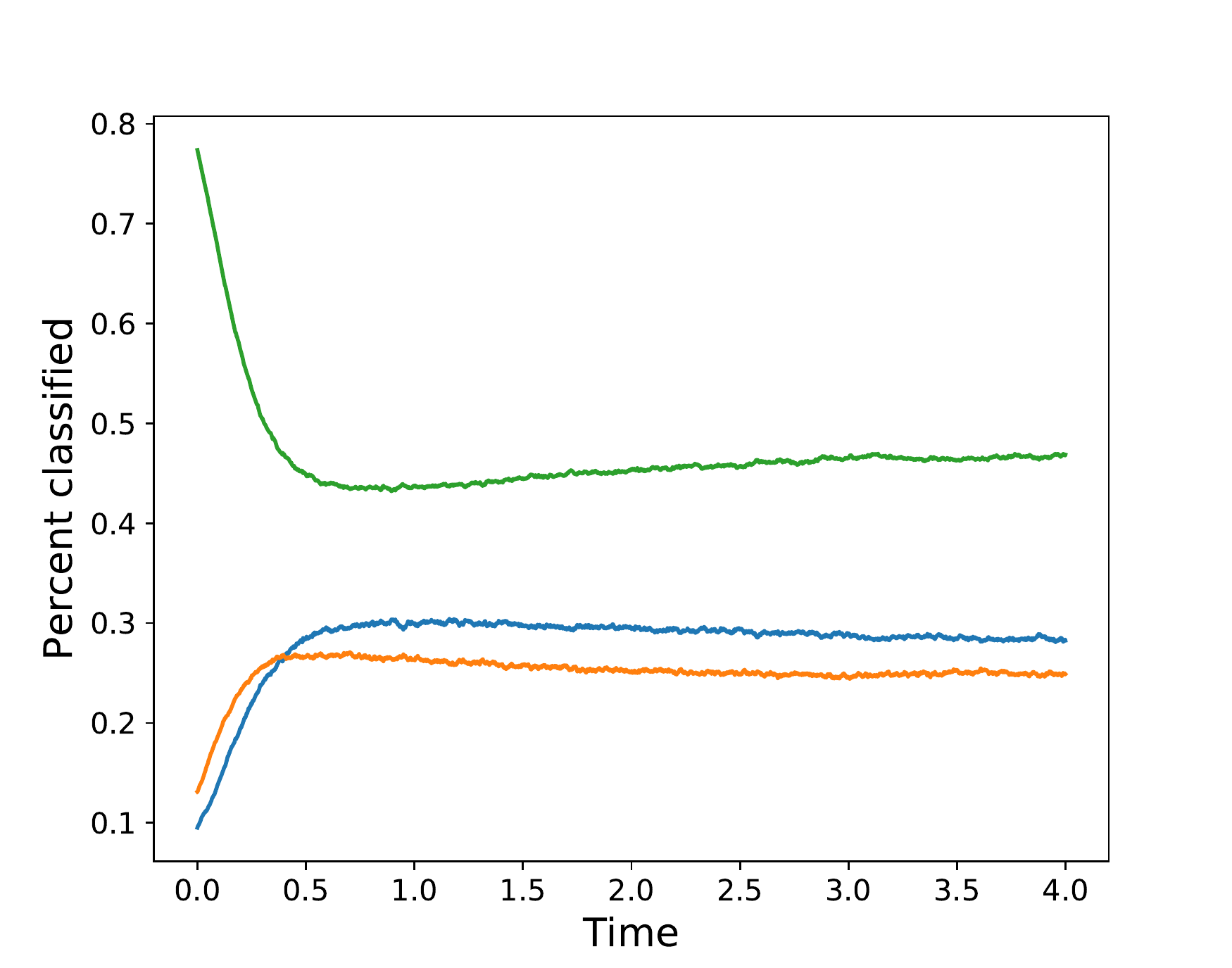}}
	} \\
\mbox{
	\subfigure[$Re=32000$, $N^2=128^2$]{\includegraphics[width=0.48\textwidth]{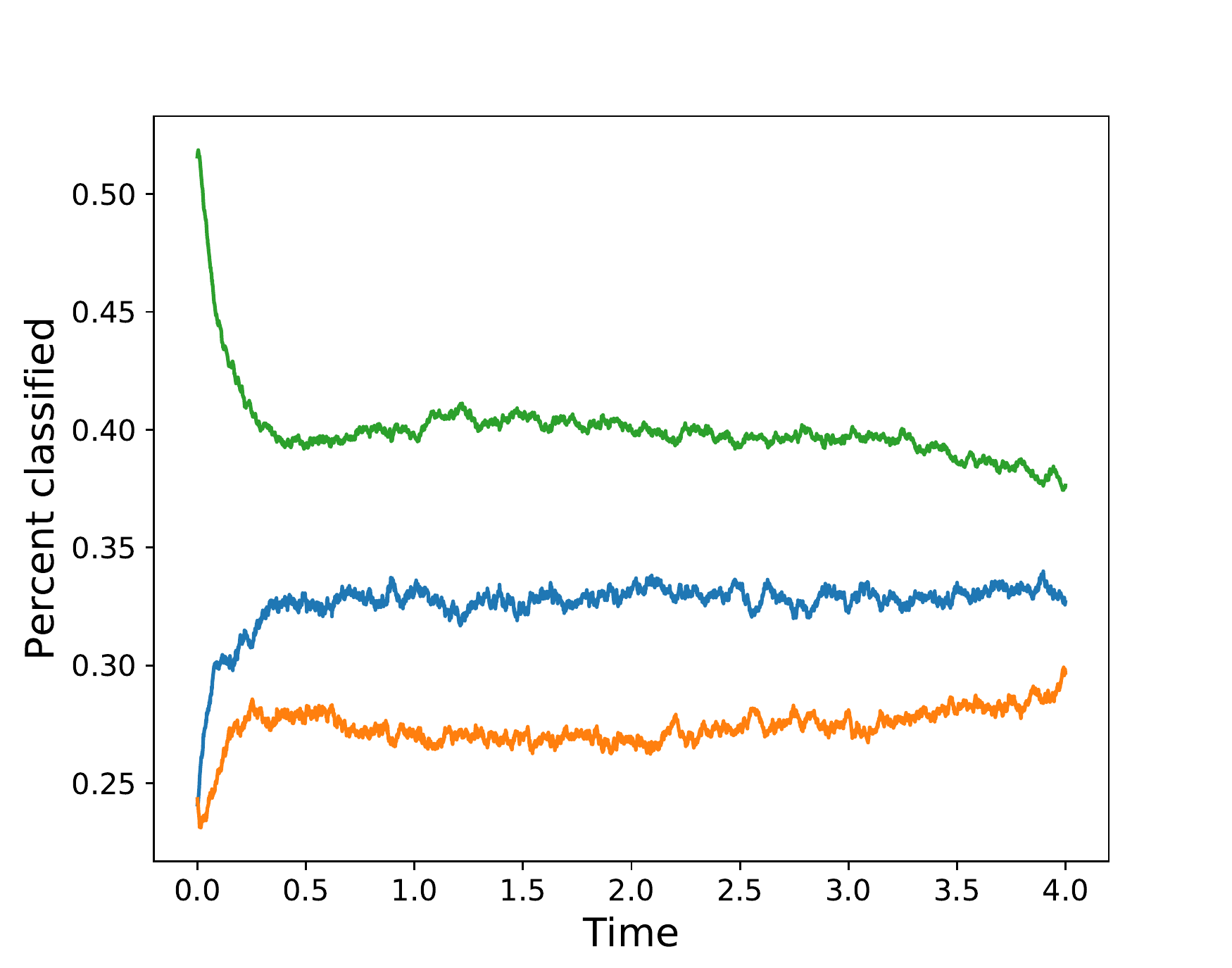}}
}
\caption{The \emph{a posteriori} classification percentages of the various modeling hypotheses for our three forward deployments. In all deployments it is observed that the utilization of AD and SM increases as the scale-separation grows and saturates for the slow decay. Noticeably, the deployment at $N^2=128^2$ necessitates a higher proportion of AD and SM classifications for improved stabilization.}
\label{Fig12b}
\end{figure}

We also determine the effect of network deployment in the presence of numerical errors as shown in table \ref{Table2} where it can be seen that a significant difference in hypothesis choices are observed. In particular, the \emph{a posteriori} deployment of the classifier is seen to utilize a greater proportion of the turbulence closure hypotheses, in comparison to the no-model ones. This may be considered as proof of the classifier detecting greater stabilization requirements due to numerical error build-up. It is observed that the AD approach shows a greater increase in deployment than SM. This may be to offset the rather large inaccuracies of the lower wavenumbers in the exceptionally dissipative SM approach. Understanding the nature of classifier adaptation in the presence of numerical errors is an interesting subject of future research that may aid in improved decision making frameworks. We complement the data in table \ref{Table2} by outlining the classification percentages of different hypotheses plotted against time for our three \emph{a posteriori} deployments in Figure \ref{Fig12b}. One may notice that the deployment of the framework at the coarser resolution of $N^2=128^2$ requires a higher degree of SM and AD classifications for successful stabilization. All experiments are seen to show a gradual increase in closure requirement as scale-separation grows with subsequent saturation in the gradual decay.

\begin{table}
\begin{center}
\begin{tabular}{ccccccc}
Time    & \multicolumn{3}{c}{\emph{A priori}} & \multicolumn{3}{c}{\emph{A posteriori}}  \\ \hline
        & AD    & SM    & No-model & AD    & SM    & No-model \\
$t = 1$ & 22.43 & 21.69 & 55.87    & 29.94 & 26.34 & 43.72 \\
$t = 2$ & 22.31 & 21.08 & 56.60    & 29.17 & 25.37 & 45.45 \\
$t = 3$ & 21.37 & 20.84 & 57.78    & 28.68 & 25.07 & 46.25 \\
$t = 4$ & 19.49 & 22.56 & 57.94    & 28.45 & 25.38 & 46.17
\end{tabular}
\caption{Classification percentages in \emph{a priori} and \emph{a posteriori}. One can see deviation from trends due to numerical error accumulation (and greater utilization of closure classifications for subsequent stabilization).}
\label{Table2}
\end{center}
\end{table}

\begin{figure}
\centering
\includegraphics[width=0.95\textwidth]{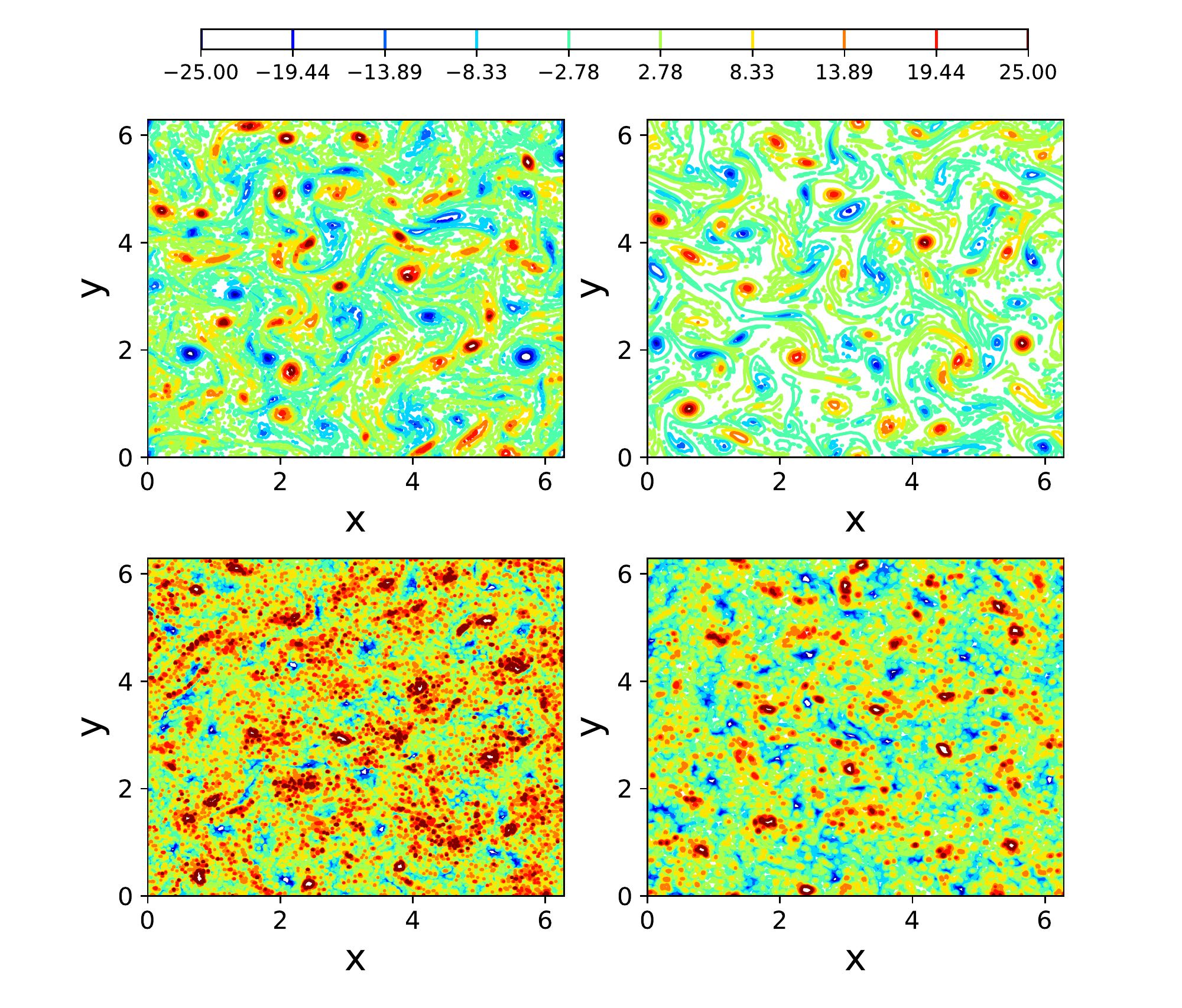}
\caption{\emph{A posteriori} contour results for $Re=32000$ with the proposed classification framework shown top-left, DS shown top-right, UNS shown bottom left and AD shown bottom right. These may be compared against FDNS contours qualitatively (in Figure \ref{Fig1}).}
\label{Fig12a}
\end{figure}

As a final qualitative analysis of our classifier, we plot \emph{a posterior} contours from forward deployments at $N^2=256^2$ and $Re=32000$. In Figure \ref{Fig12a}, vorticity contours from the ML, DS, AD and UNS simulations are shown to assess the stabilization effect of the different frameworks. The classifier can be seen to stabilize high-wavenumber noise adequately, in a manner similar to DS as previous statistics have reflected. The AD approach may be observed to be contaminated with noise that may potentially be harmful for long-time integration. 

\subsection{Model blending}
\label{model_blending}

In this section, we deploy our learning in a different manner by utilizing their outputs (i.e., the conditional probabilities of each hypothesis) as a pre-multiplier of the prediction of each modeling hypothesis. We utilize this formulation instead of the direct prediction of sub-grid contribution coefficients by observing that a greater degree of stability is imparted to the flow-evolution. Indeed, direct regression with sub-grid quantities has been seen to require \emph{a posteriori} post-processing for stability \citep{maulik2018deconvolution,maulik2019subgrid} due to energy accummulation in the super-grid (when negative eddy-viscosities are predicted effectively). We recognize (as a limitation), that the utilization of a conditional-probability outputs to linearly combine turbulence modeling predictions from different hypotheses digresses from the core idea of a categorical cross-entropy error minimization. However, as results shall show, the proposed method acts as an effective instrument for blending models in \emph{a posteriori} with the requirement of any truncation for stability. We would also like to emphasize here that the same learning is applicable for both classification and blending. We perform a similar set of assessments as outlined in Section~\ref{model_class}. 

\begin{figure}
\centering
\mbox{
\subfigure{\includegraphics[width=0.48\textwidth]{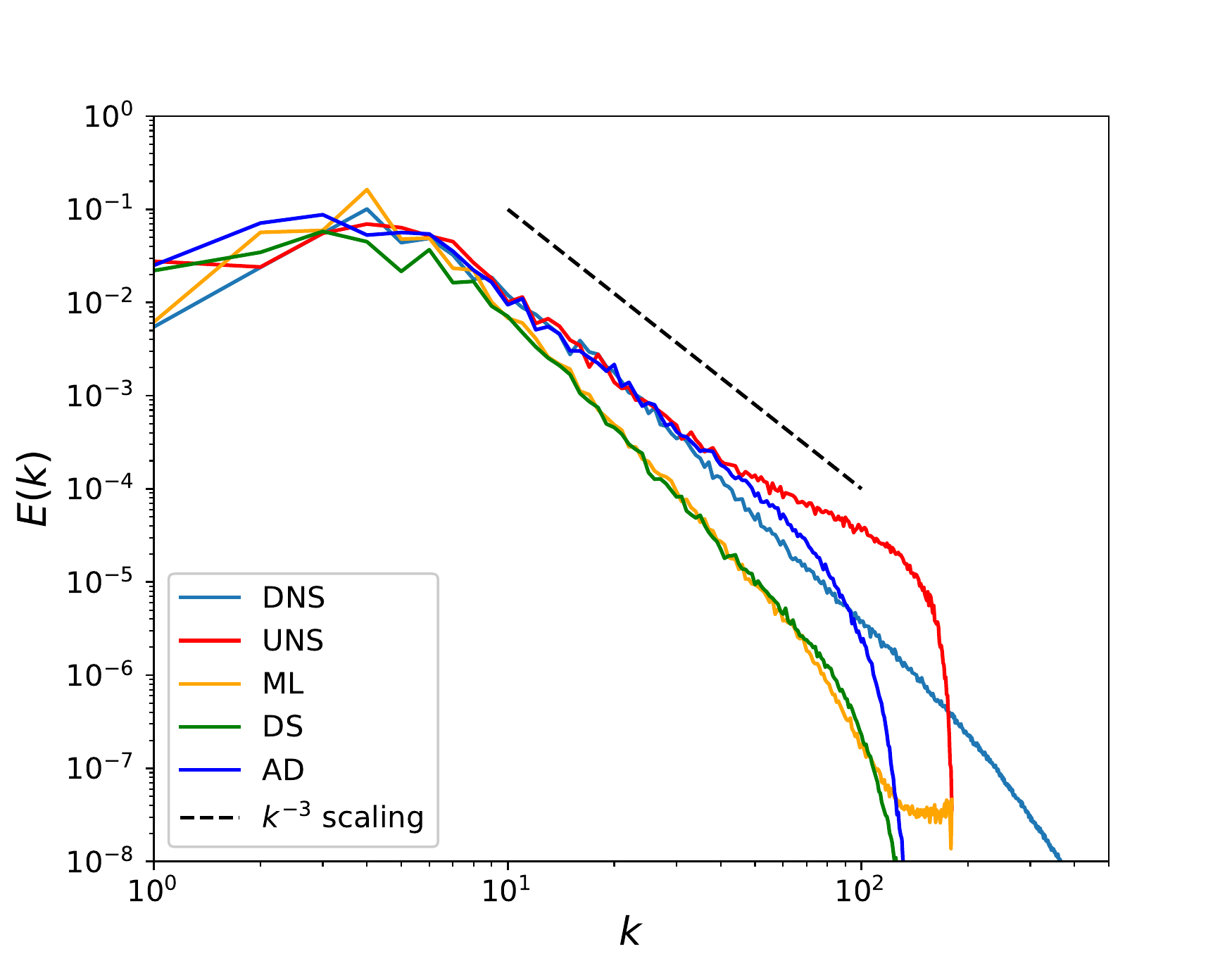}}
\subfigure{\includegraphics[width=0.48\textwidth]{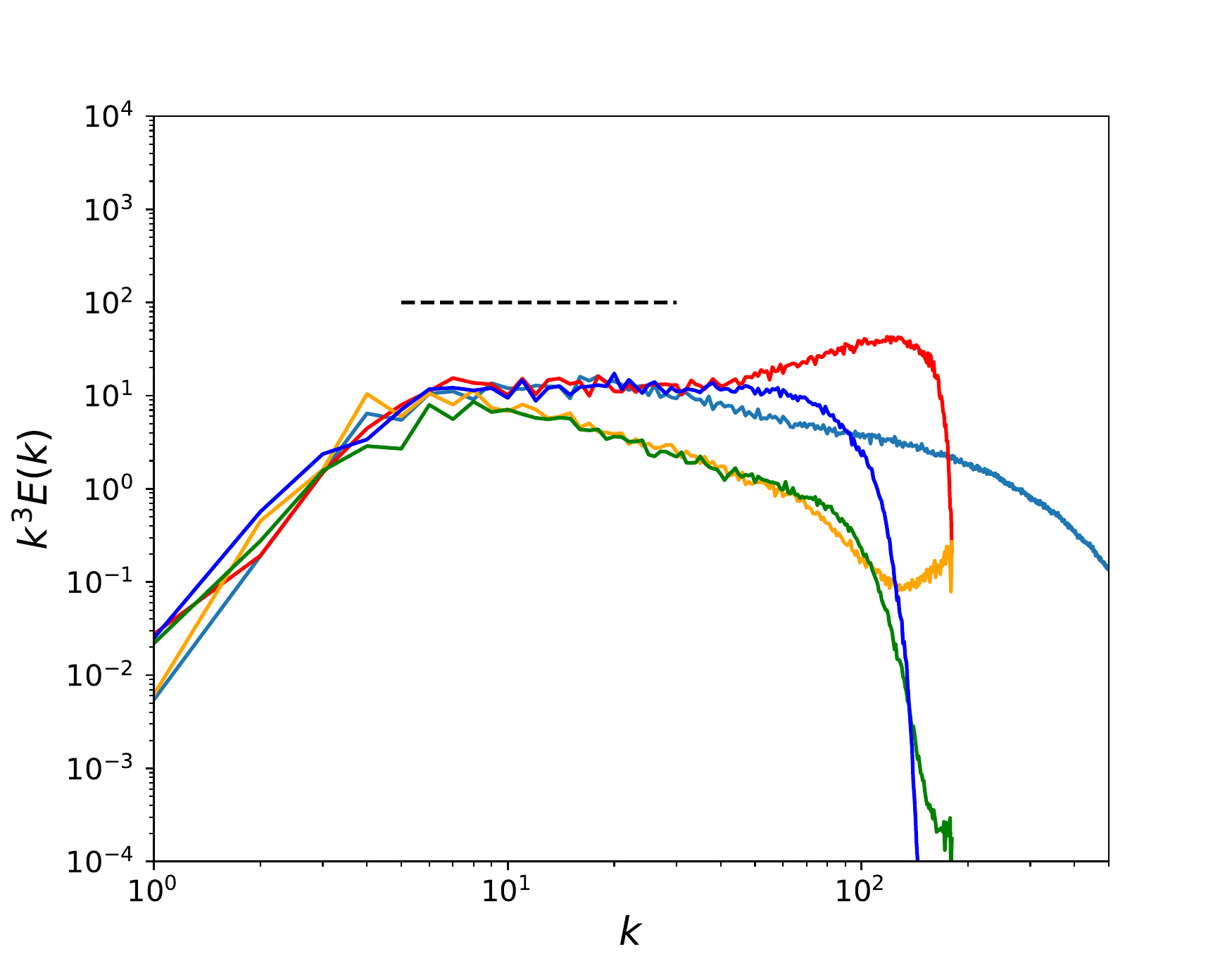}}
}
\caption{\emph{A posteriori} kinetic-energy spectra (left) and compensated kinetic-energy spectra (right) for $Re=32000$ at $t=4$ and at $N^2=256^2$ degrees of freedom. The proposed framework (deployed as a model blending mechanism) behaves similar to the DS approach at the inertial wavenumbers. We remind the reader that the blending is dynamic between AD and SM.}
\label{Fig13}
\end{figure}

\begin{figure}
\centering
\mbox{
\subfigure{\includegraphics[width=0.48\textwidth]{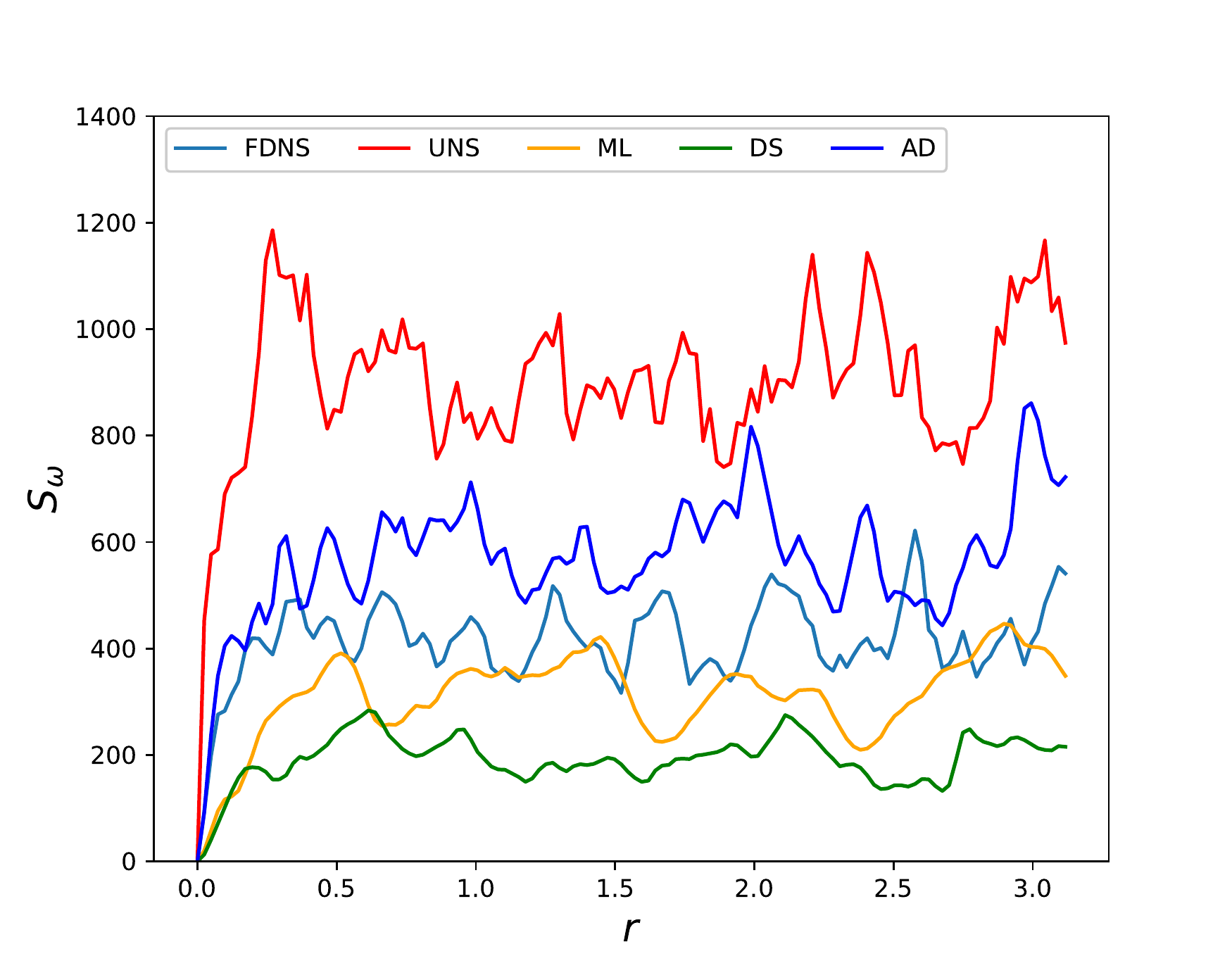}}
\subfigure{\includegraphics[width=0.48\textwidth]{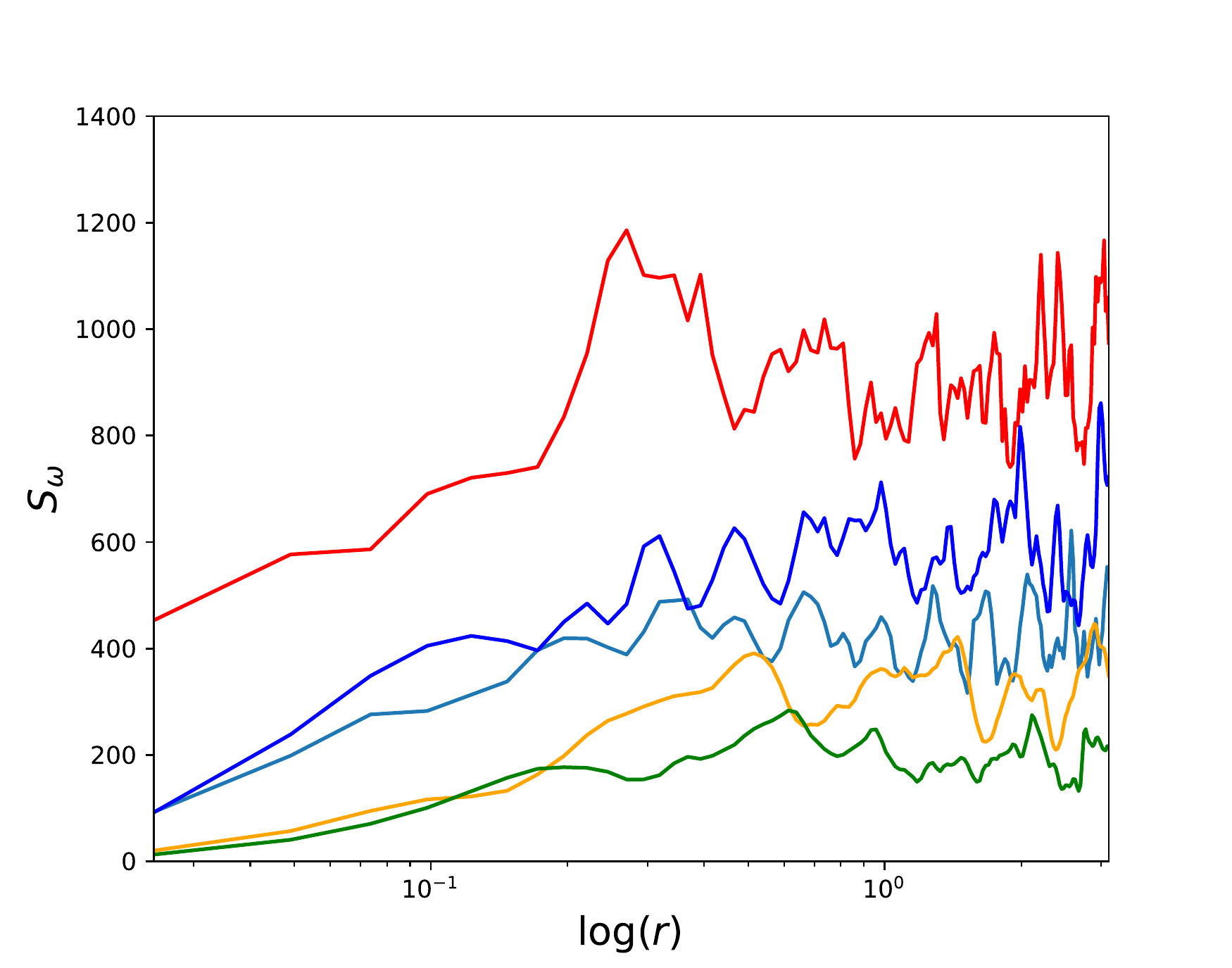}}
}
\caption{\emph{A posteriori} vorticity structure functions plotted against $\textbf{r}$ (left) and $\log(\textbf{r})$ (right) for $Re=32000$ at $t=4$ and at $N^2=256^2$ degrees of freedom. It is observed that solely AD performs better in the near-region whereas the proposed blending (once again) behaves similar to the DS approach. We remind the reader that the blending is dynamic between AD and SM.}
\label{Fig14}
\end{figure}

\begin{figure}
\centering
\mbox{
\subfigure{\includegraphics[width=0.48\textwidth]{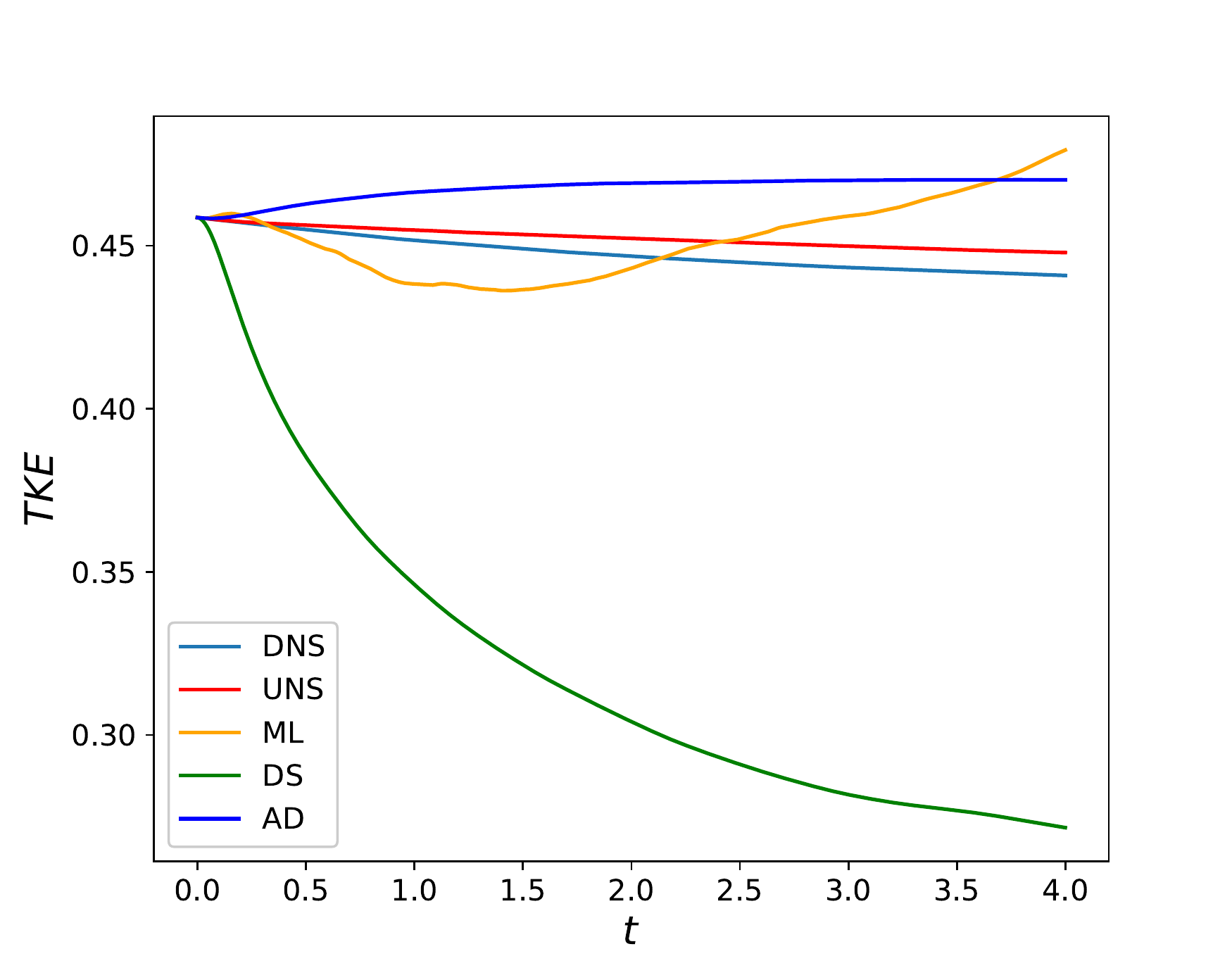}}
\subfigure{\includegraphics[width=0.48\textwidth]{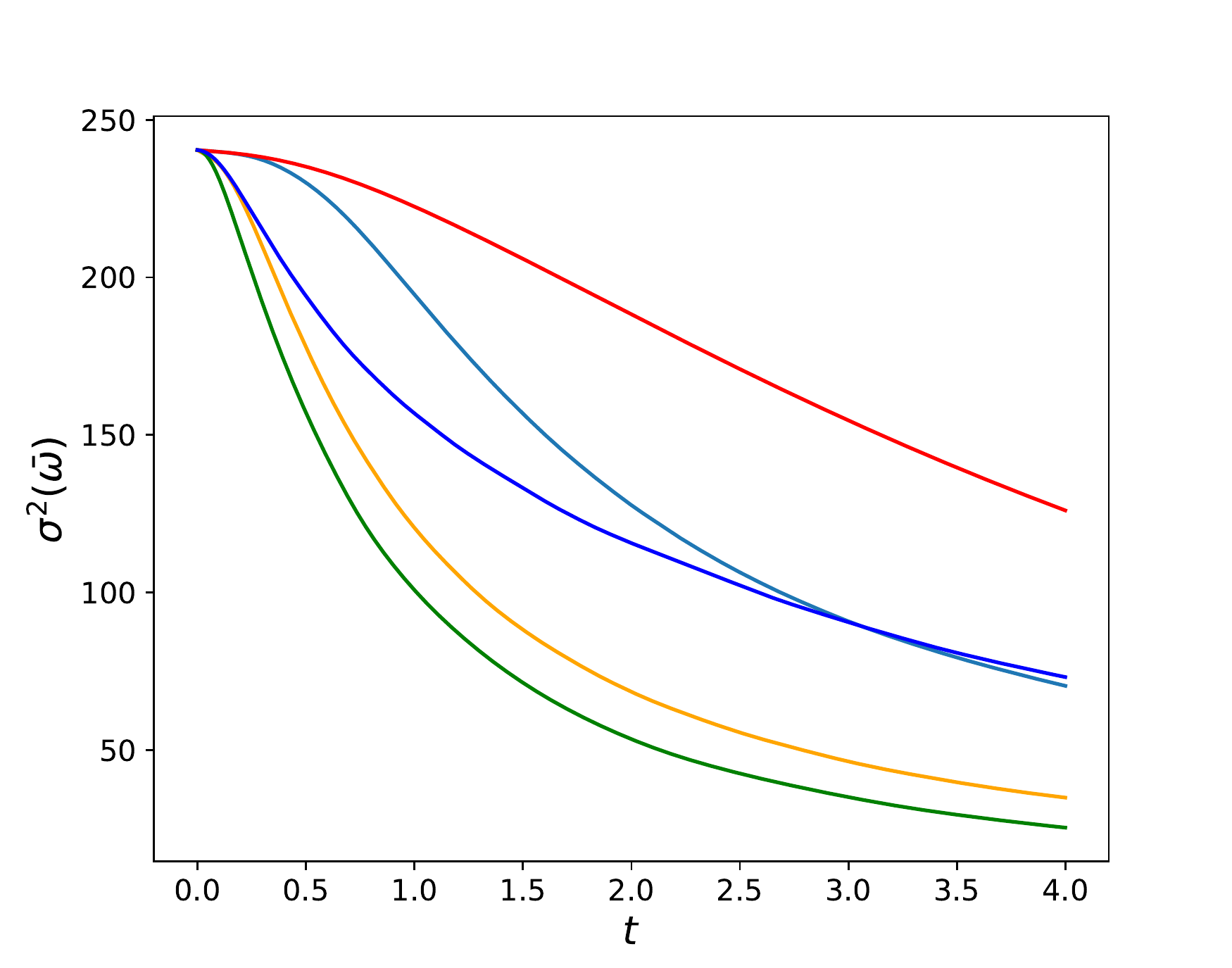}}
}
\caption{Time-histories for turbulent kinetic energy (left) and vorticity variance (right) for $Re=32000$ at $N^2=256^2$ degrees of freedom. The proposed blending technique shows a varying TKE capture behavior due to its adaptive dissipation. Note that the blending is dynamic between AD and SM.}
\label{Fig15}
\end{figure}

Figure \ref{Fig13} shows the performance of the blending formulation for a Reynolds number of 32000 and at $t=4$ with $N^2=256^2$ degrees-of-freedom with kinetic energy spectra. It is observed that the proposed procedure recovers a dissipative behavior that is very similar to the DS approach. It does this by balancing the coefficients of the AD and SM predictions which adapt to the dynamic dissipation requirement of the flow. Overall, it is observed that the framework behaves in a similar manner to the classifier presented previously with dissipation preventing the accummulation of high-wavenumber errors but causing a mismatch in the inertial range spectra capture. However, the dissipation is dynamic and it prevents the overwhelming damping of the SM deployment by balancing with the AD predictions adaptively. This is reflected in Figure \ref{Fig14} as well where the vorticity structure functions once again show that the AD method is more accurate at lower values of $\textbf{r}$ but the blending allows for a prediction akin to the DS technique. Figure \ref{Fig15} shows the time-histories of the TKE and the $\sigma^2 (\bar{\omega})$ for the proposed framework compared to DS, AD and UNS. The vorticity-variance shows a trend close to the DS approach as expected but the TKE trends are once again not uniform. 

\begin{figure}
\centering
\mbox{
\subfigure{\includegraphics[width=0.48\textwidth]{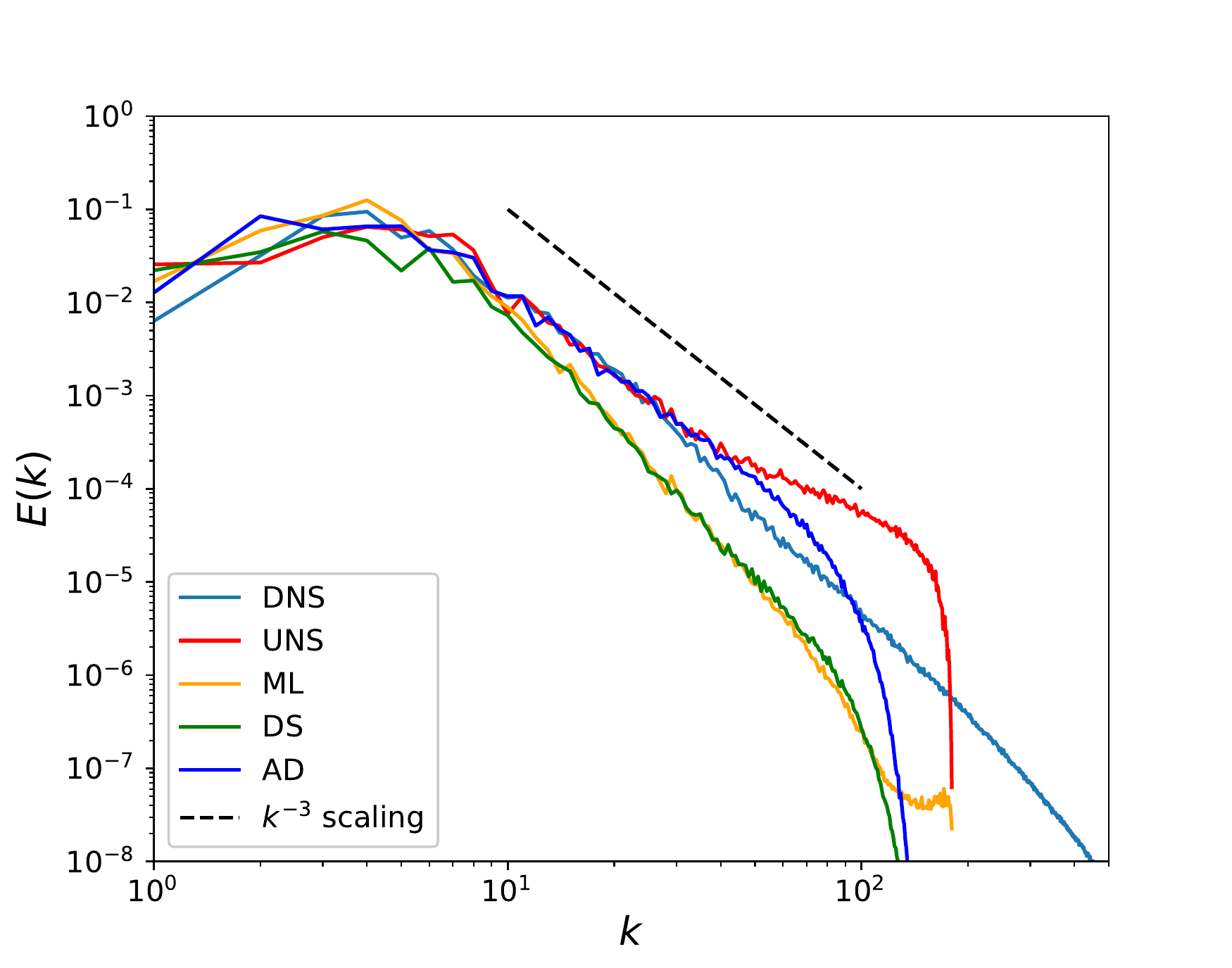}}
\subfigure{\includegraphics[width=0.48\textwidth]{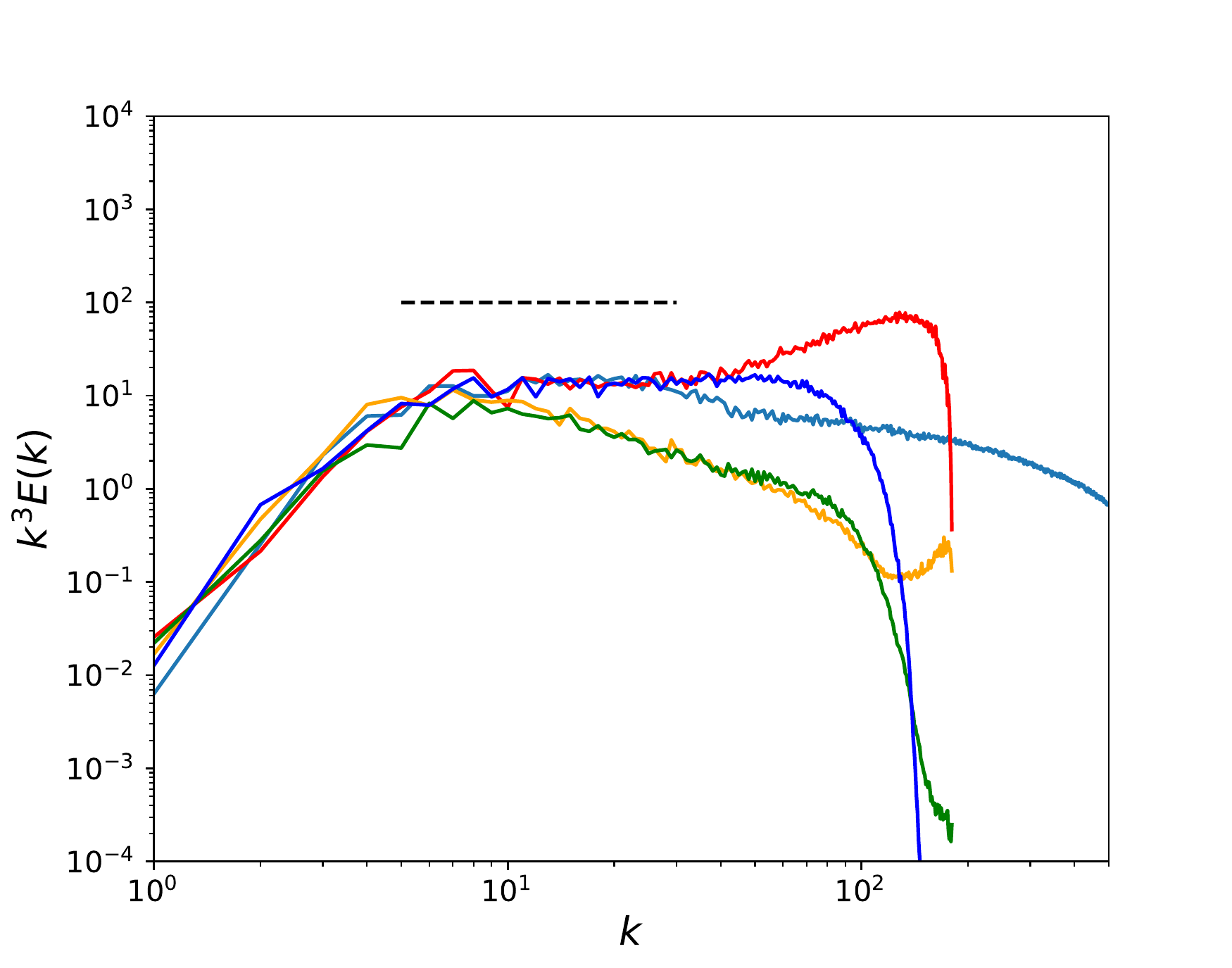}}
}
\caption{\emph{A posteriori} kinetic-energy spectra (left) and compensated kinetic-energy spectra (right) for $Re=64000$ at $t=4$ and at $N^2=256^2$ degrees of freedom. The proposed framework (deployed as a model blending mechanism) behaves similar to the DS approach at the inertial wavenumbers. Note that the blending is dynamic between AD and SM and training is performed using $Re=32000$ data alone.}
\label{Fig16}
\end{figure}

\begin{figure}
\centering
\mbox{
\subfigure{\includegraphics[width=0.48\textwidth]{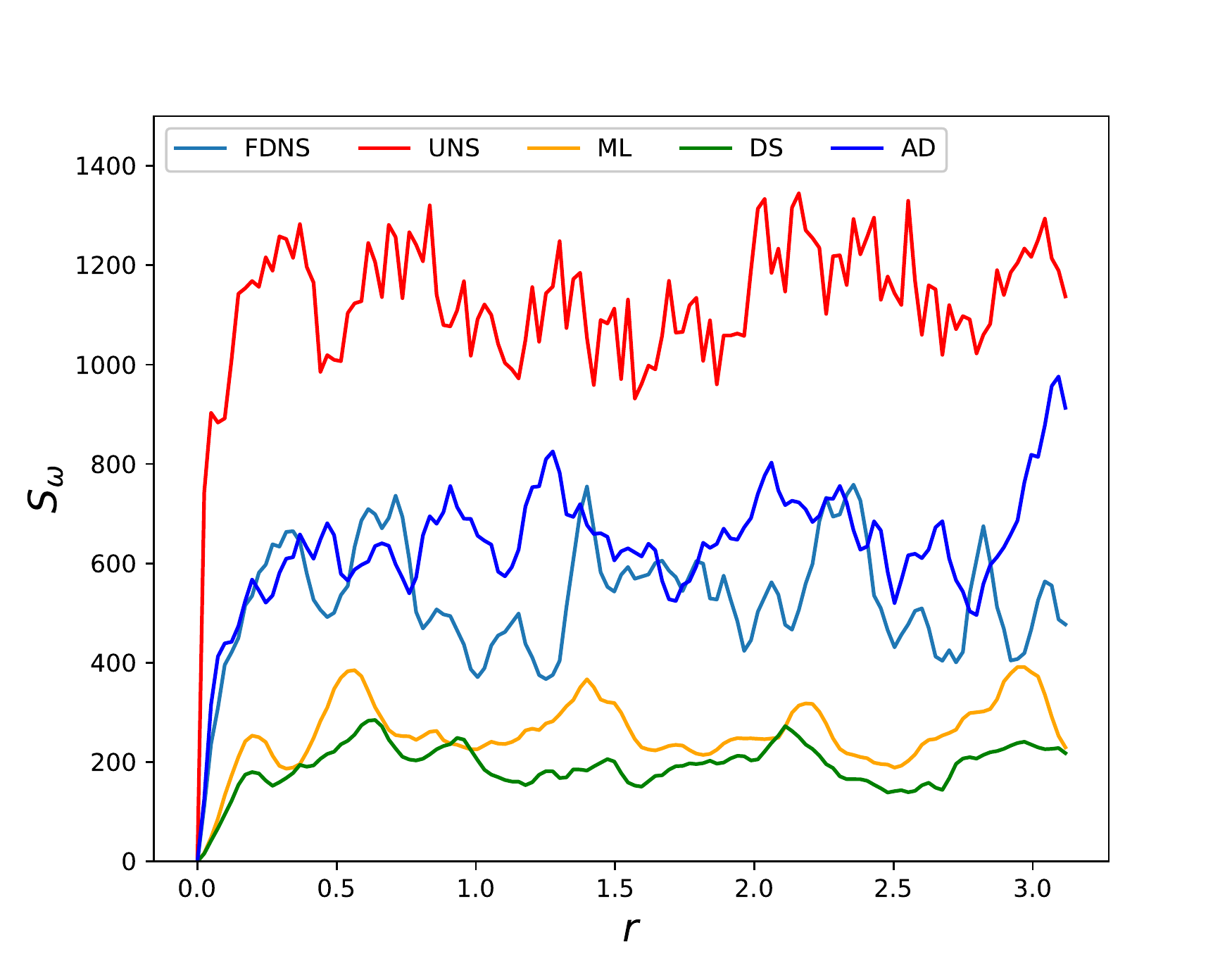}}
\subfigure{\includegraphics[width=0.48\textwidth]{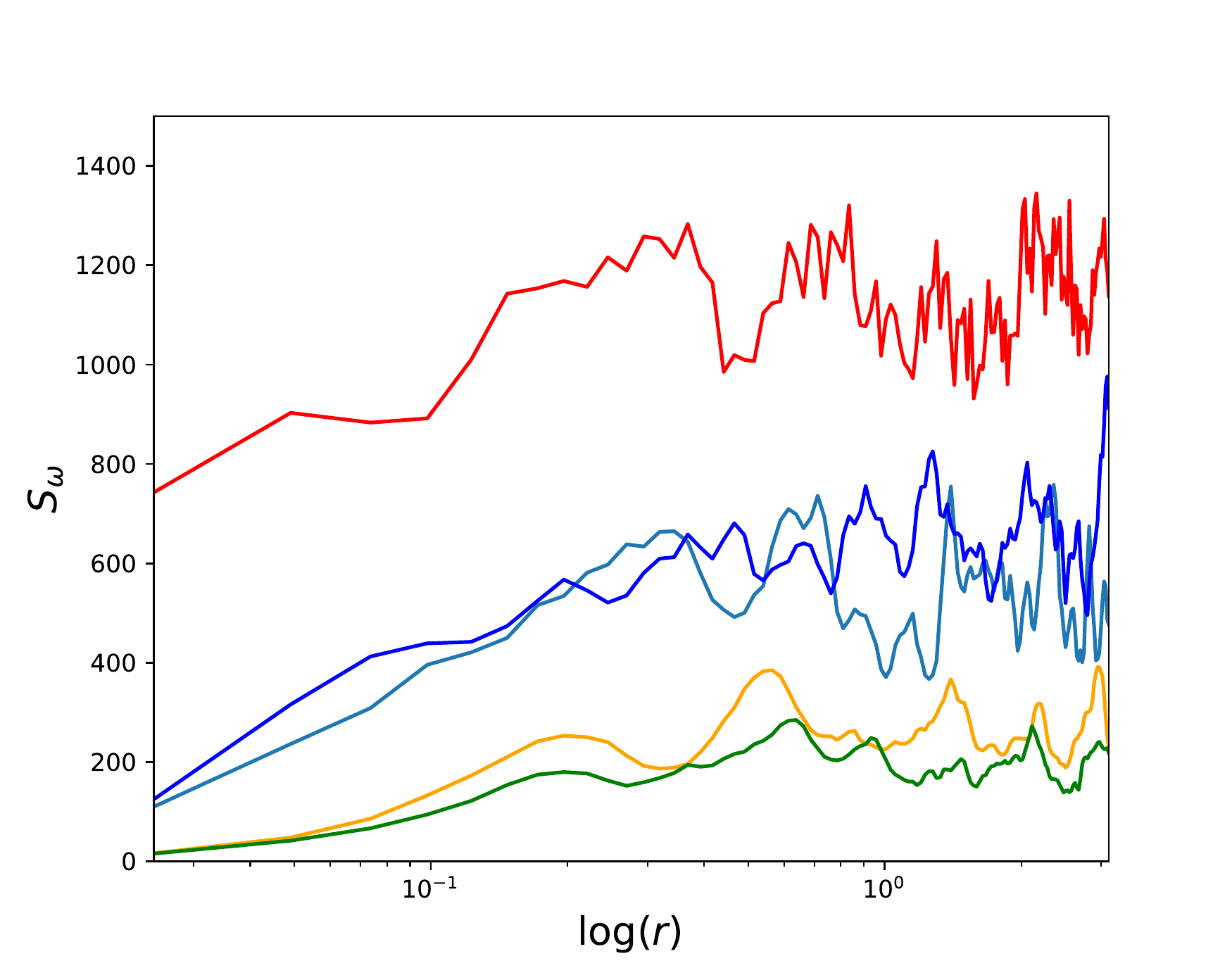}}
}
\caption{\emph{A posteriori} vorticity structure functions plotted against $\textbf{r}$ (left) and $\log(\textbf{r})$ (right) for $Re=64000$ at $t=4$ and at $N^2=256^2$ degrees of freedom. It is observed that solely AD performs better in the near-region whereas the proposed blending (once again) behaves similar to the DS approach. We remind the reader that the blending is dynamic between AD and SM and training is performed using $Re=32000$ data alone.}
\label{Fig17}
\end{figure}

\begin{figure}
\centering
\mbox{
\subfigure{\includegraphics[width=0.48\textwidth]{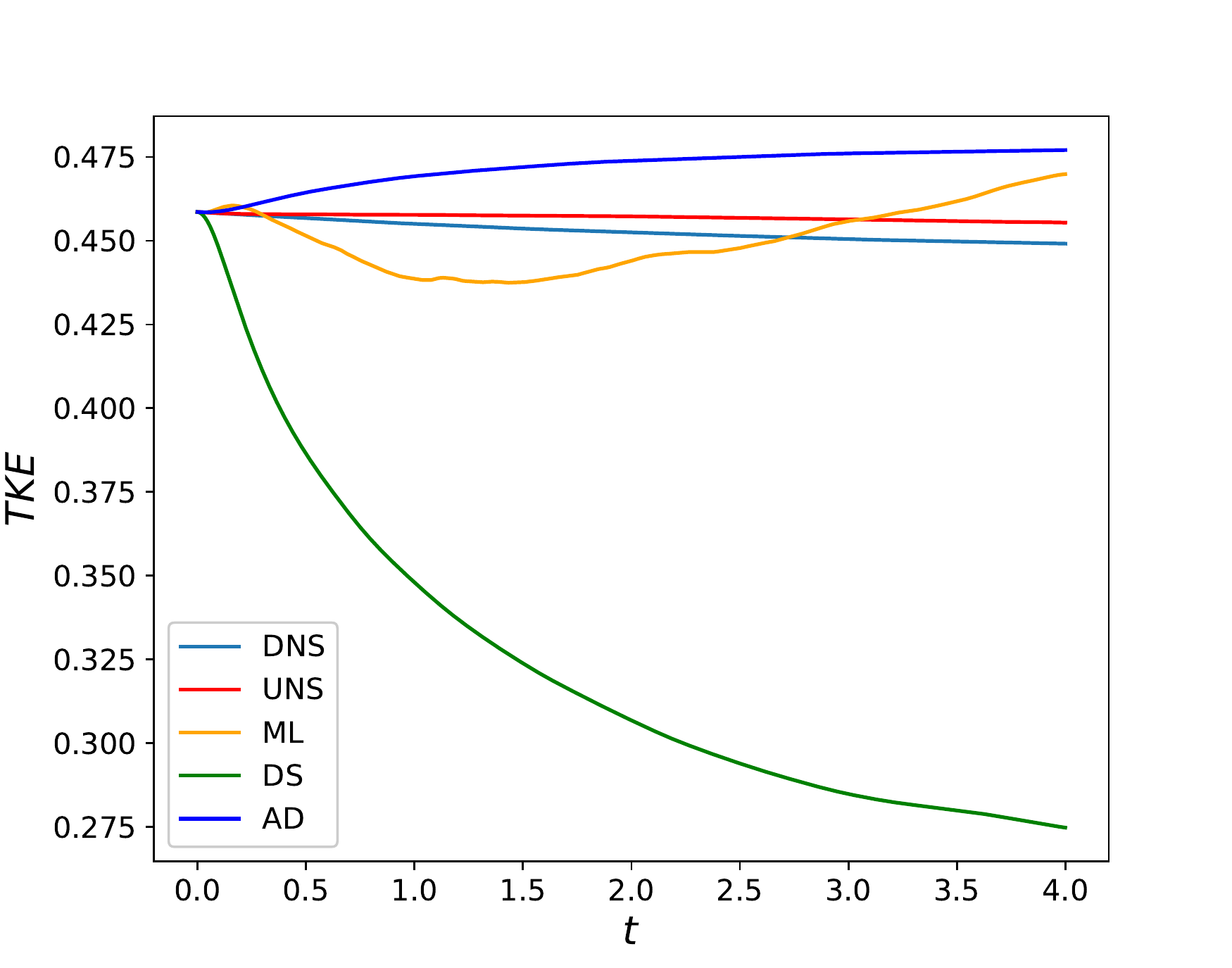}}
\subfigure{\includegraphics[width=0.48\textwidth]{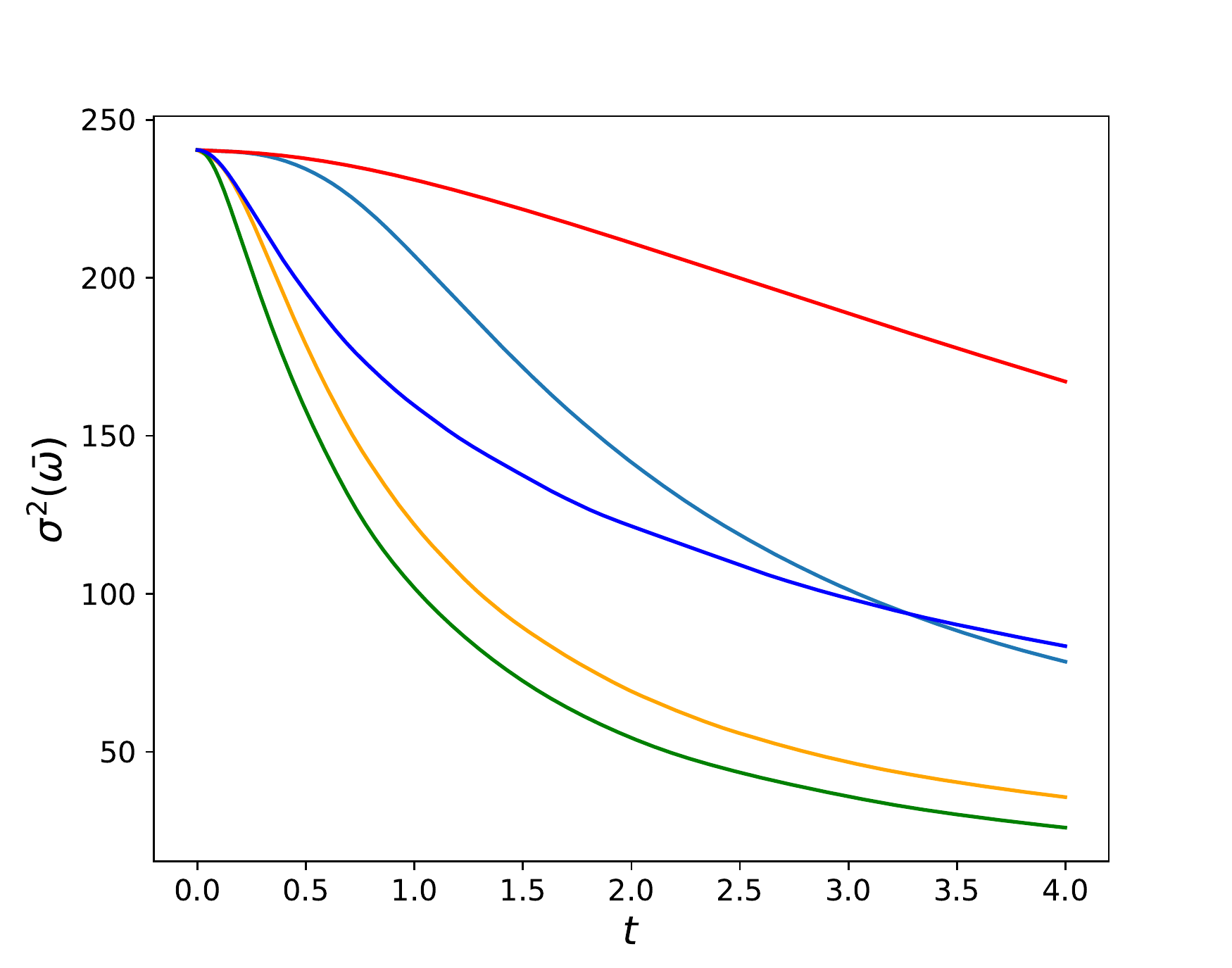}}
}
\caption{Time-histories for turbulent kinetic energy (left) and vorticity variance (right) for $Re=64000$ at $N^2=256^2$ degrees of freedom. We remind the reader that the blending is dynamic between AD and SM and training is performed using $Re=32000$ data alone.}
\label{Fig18}
\end{figure}

\begin{figure}
\centering
\mbox{
\subfigure{\includegraphics[width=0.48\textwidth]{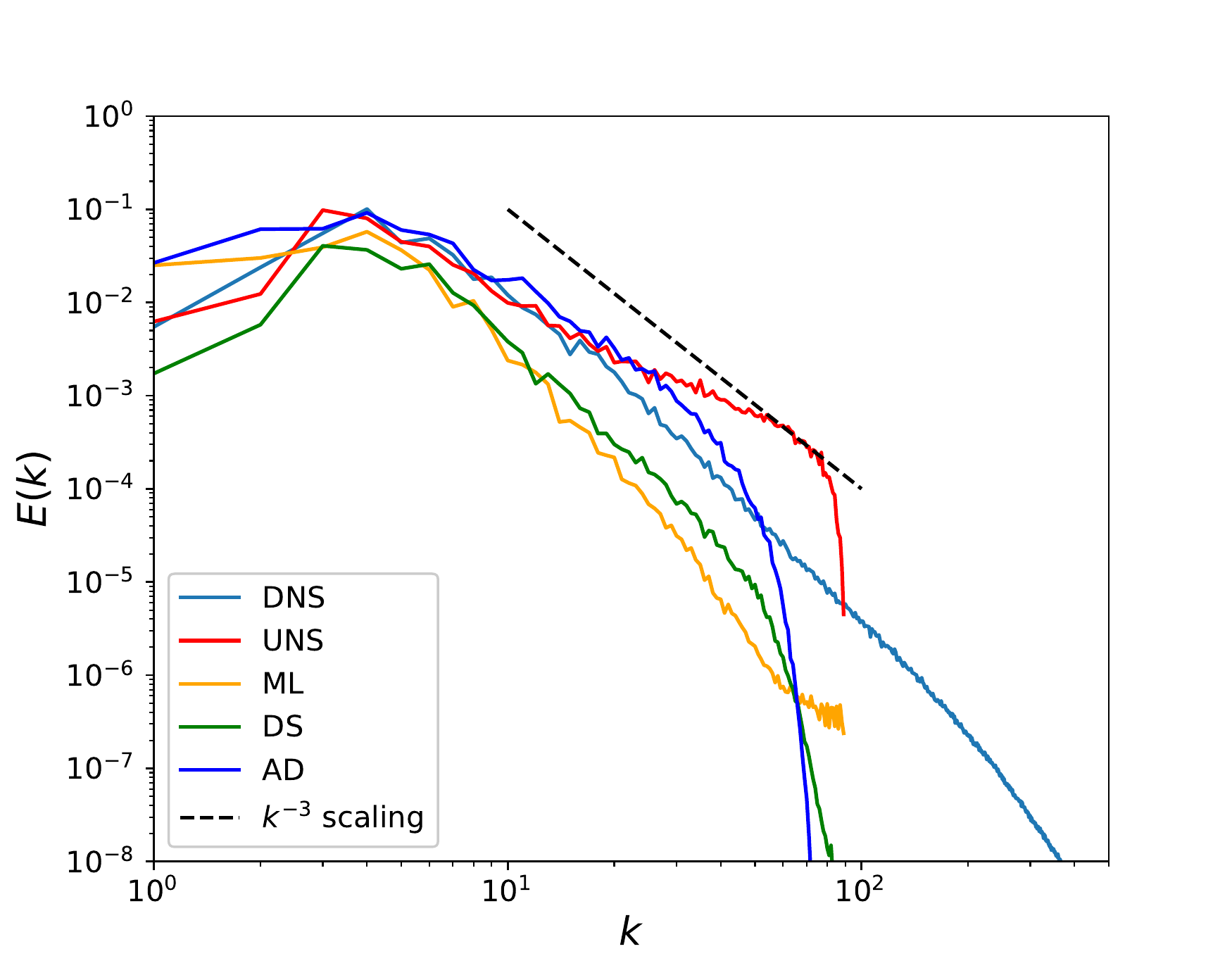}}
\subfigure{\includegraphics[width=0.48\textwidth]{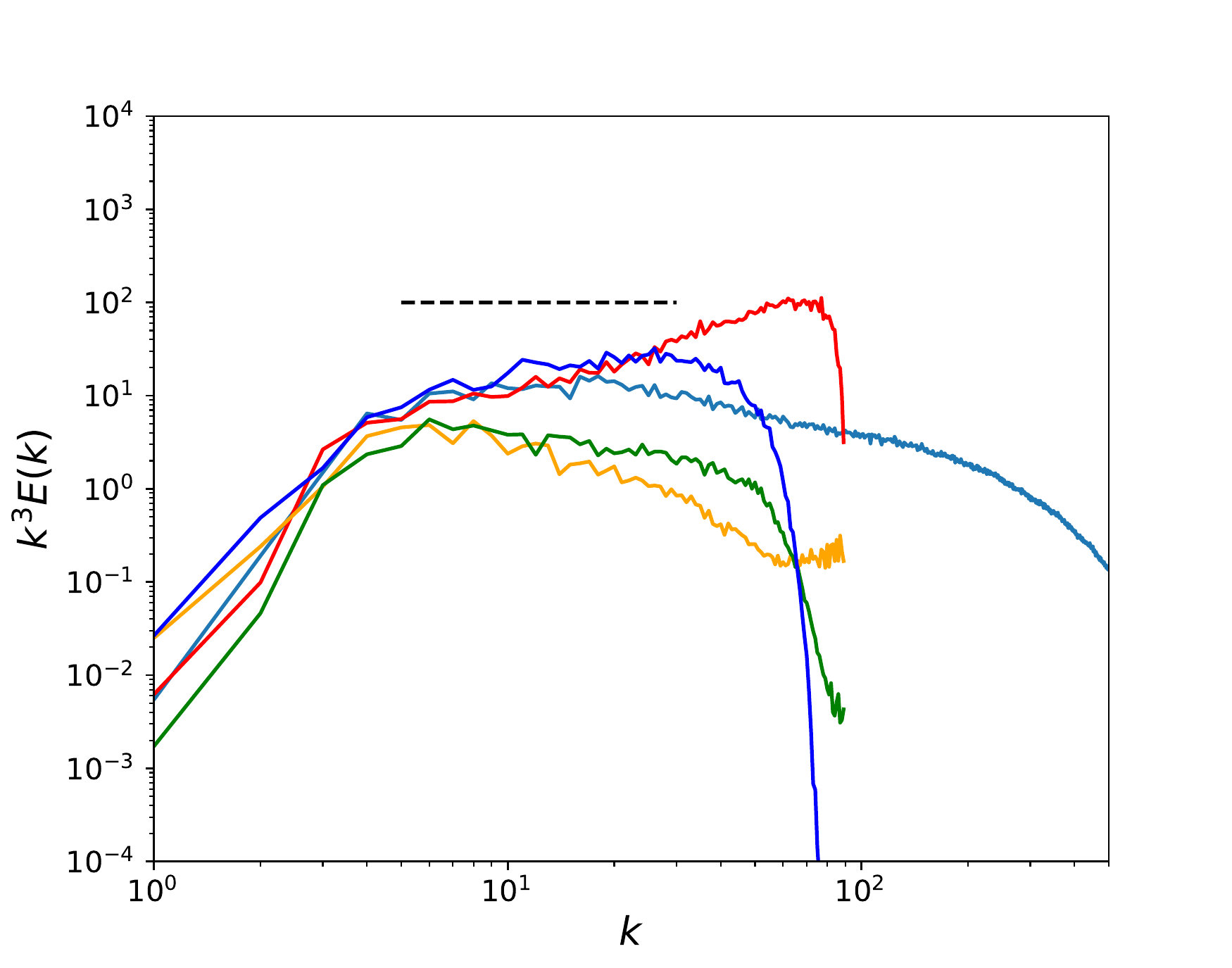}}
}
\caption{\emph{A posteriori} kinetic-energy spectra (left) and compensated kinetic-energy spectra (right) for $Re=32000$ at $t=4$ and at $N^2=128^2$ degrees of freedom. The proposed framework (deployed as a model blending mechanism) behaves similar to the DS approach at the inertial wavenumbers. We remind the reader that the blending is dynamic between AD and SM and training is performed using $Re=32000$ and $N^2=256^2$ data alone.}
\label{Fig19}
\end{figure}

\begin{figure}
\centering
\mbox{
\subfigure{\includegraphics[width=0.48\textwidth]{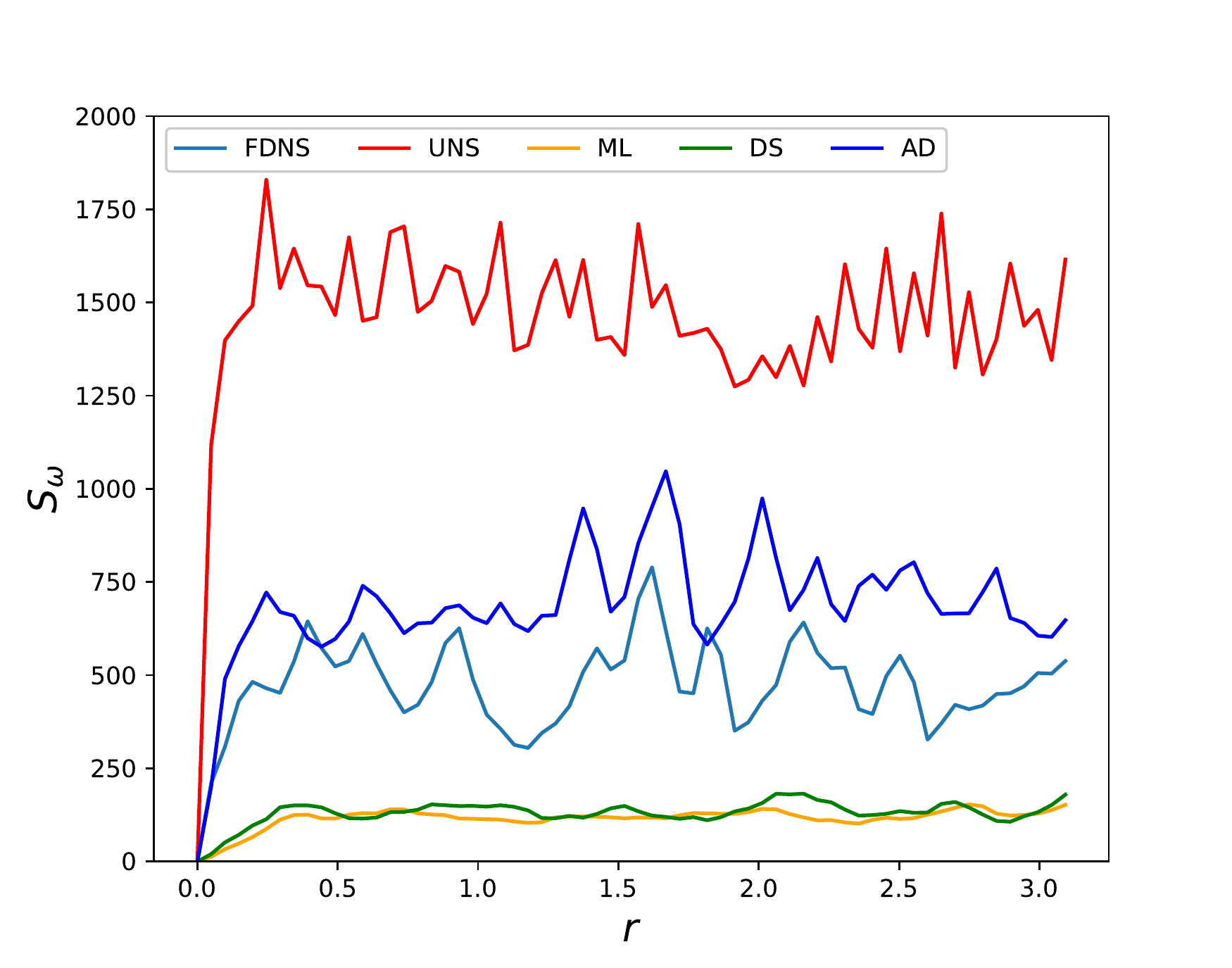}}
\subfigure{\includegraphics[width=0.48\textwidth]{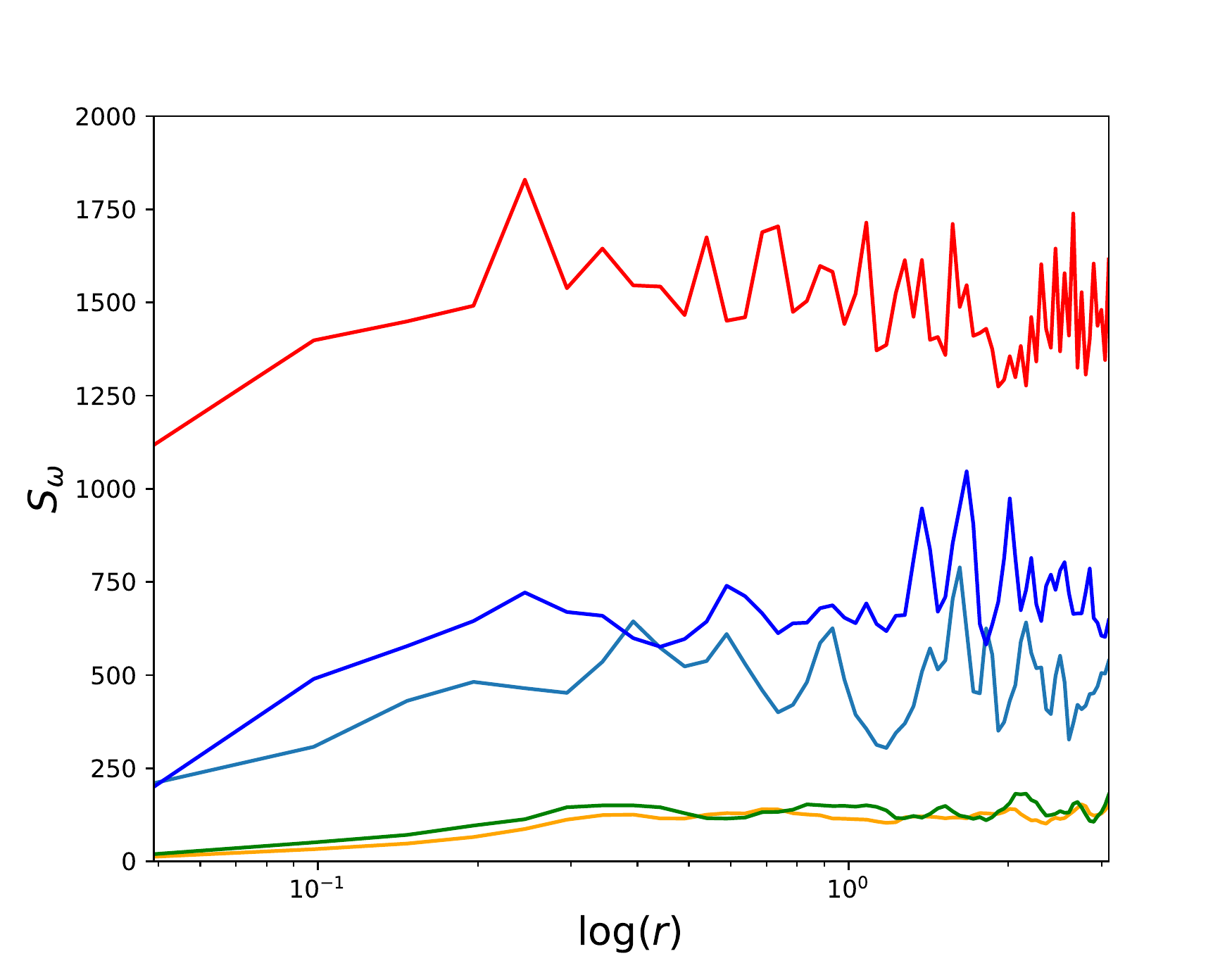}}
}
\caption{\emph{A posteriori} vorticity structure functions plotted against $\textbf{r}$ (left) and $\log(\textbf{r})$ (right) for $Re=32000$ at $t=4$ and at $N^2=128^2$ degrees of freedom. It is observed that solely AD performs better in the near-region whereas the proposed blending (once again) behaves similar to the DS approach. We remind the reader that the blending is dynamic between AD and SM and training is performed using $Re=32000$ and $N^2=256^2$ data alone.}
\label{Fig20}
\end{figure}

\begin{figure}
\centering
\mbox{
\subfigure{\includegraphics[width=0.48\textwidth]{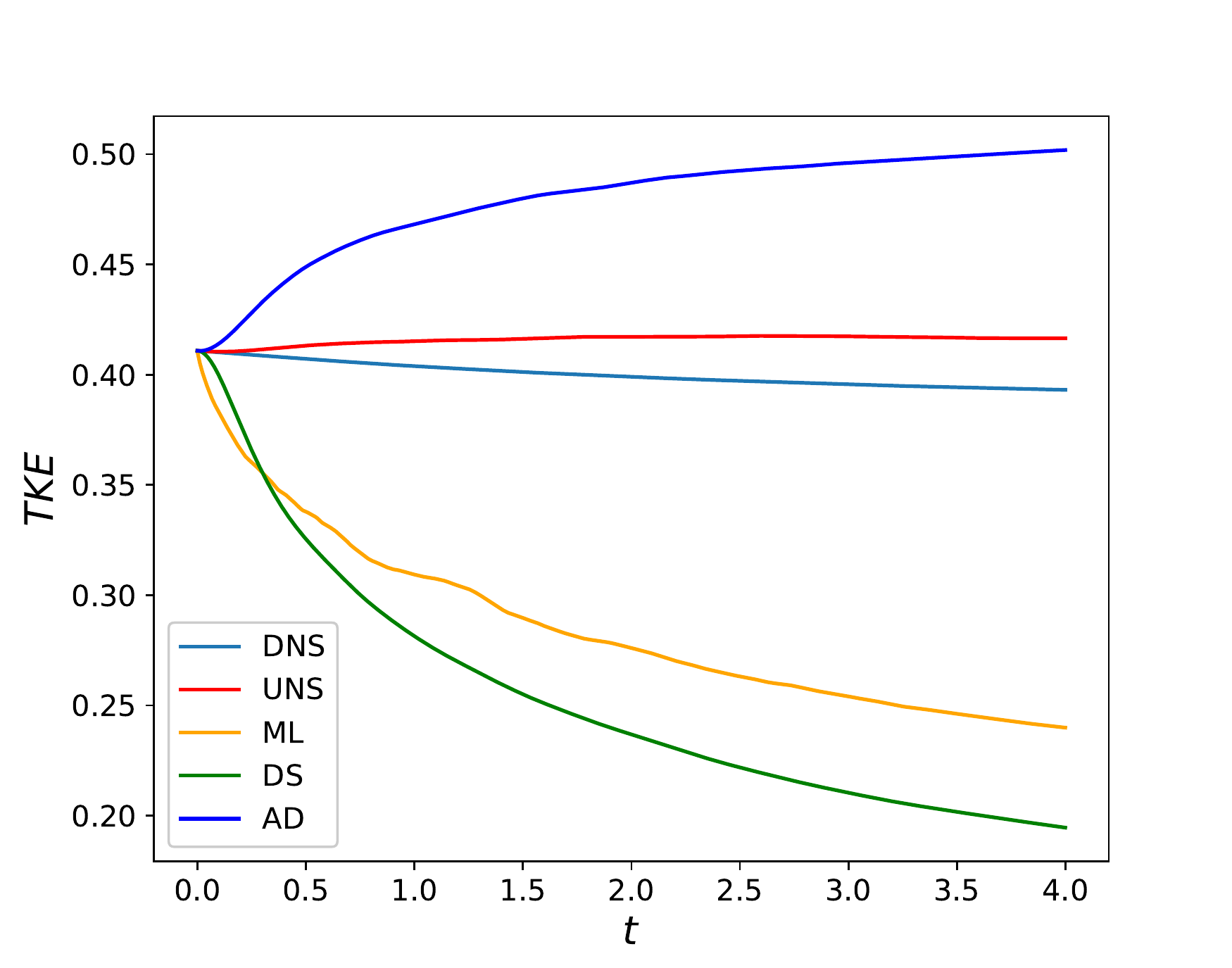}}
\subfigure{\includegraphics[width=0.48\textwidth]{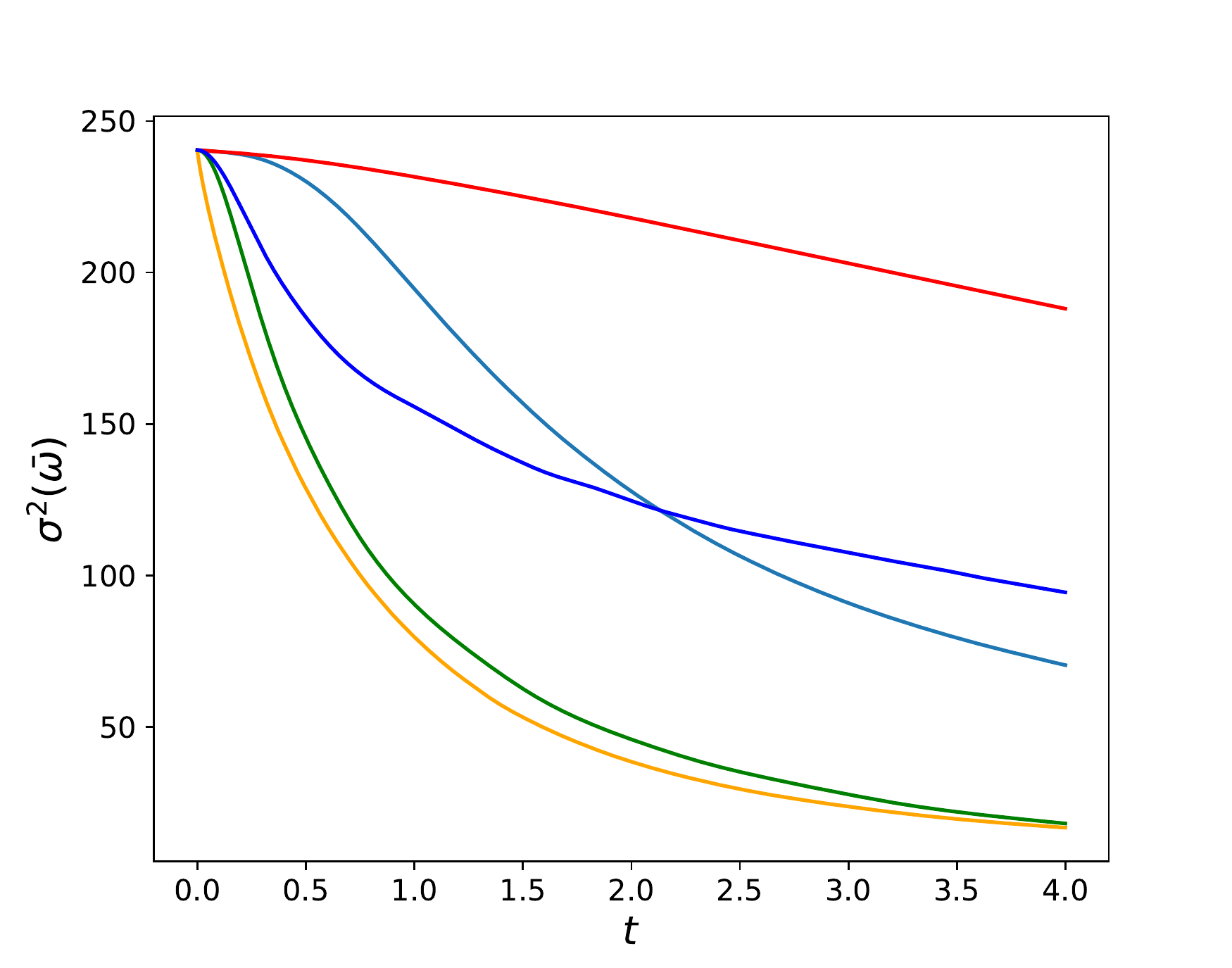}}
}
\caption{Time-histories for turbulent kinetic energy (left) and vorticity variance (right) for $Re=32000$ at $N^2=128^2$ degrees of freedom. The proposed blending technique behaves more dissipatively due to the reduced grid-support. We remind the reader that the blending is dynamic between AD and SM and training is performed using $Re=32000$ and $N^2=256^2$ data alone.}
\label{Fig21}
\end{figure}

In a fashion similar to that employed in section \ref{model_class}, we deploy assessments of the blending method to out-of-training predictions for validation. We start with an \emph{a poseirori} deployment at $Re=64000$ and $N^2=256^2$ degrees-of-freedom and observe that the learning is sufficiently generalizable. This is observed from Figure \ref{Fig16} where kinetic energy spectra show an aligned prediction to the previous test-case. Vorticity structure functions and time-histories, shown in Figures \ref{Fig17} and \ref{Fig18} respectively show a similar behavior to the one observed for $Re=32000$. This implies that the learning, whether utilized as a classifier or a blending mechanism, is generalizable. We also deploy the blending framework at a different degree-of-freedom ($N^2=128^2$) to assess it is stable to a slightly different grid support and trends similar to the classifier are observed wherein the framework focuses on dissipation to stabilize the higher wavenumbers in contrast with AD. This is observed in Figure \ref{Fig19} for kinetic energy spectra and Fig \ref{Fig20} for the vorticity structure functions. Figure \ref{Fig21} shows time-series quantities for this test-case with both TKE and vorticity-variance trends resembling the DS method closely. This also echoes with the performance of the classifier where a coarser-grid resolution led to a performance that was observed to be biased towards the eddy-viscosity hypothesis. However, further studies are necessary to quantify how the model orients itself to compensate for loss of grid-resolution or anisotropies in the flow configuration in \emph{a posteriori} deployment.

\begin{figure}
\centering
\includegraphics[width=0.95\textwidth]{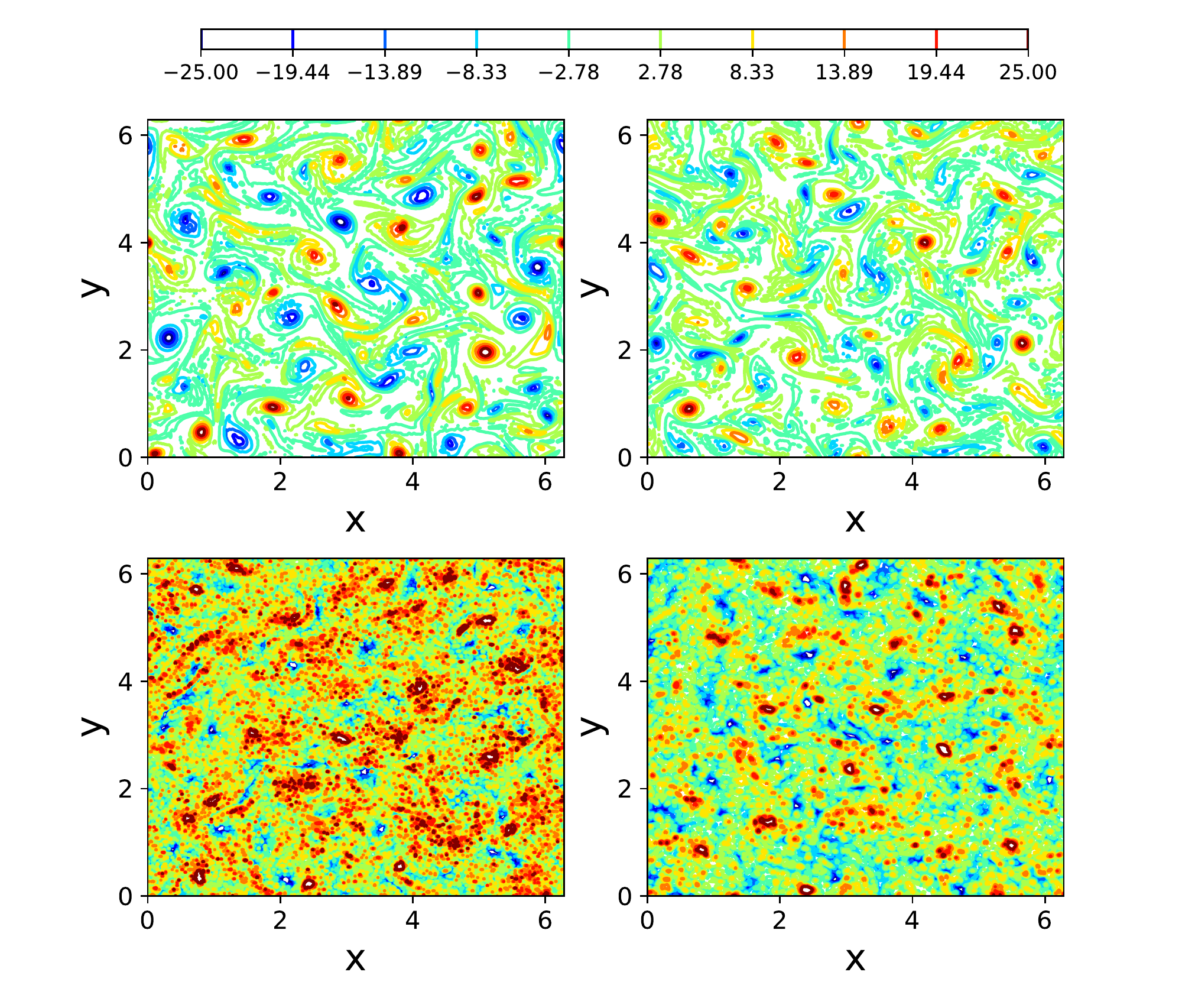}
\caption{\emph{A posteriori} contour results for $Re=32000$ with the proposed blending framework shown top-left, DS shown top-right, UNS shown bottom left and AD shown bottom right. These may be compared against FDNS contours qualitatively (in Figure \ref{Fig1}).}
\label{Fig21a}
\end{figure}

To conclude this section we show qualitative results from the vorticty contours at the final time of the numerical experiments for our proposed framework and their benchmark counterparts in Figure \ref{Fig21a}. This examination gives us an intuition of the stabilization effect of the proposed framework and it is seen that the predictions are very closely aligned with the DS results. This, once again, validates our dynamic dissipation hypothesis. 

\section{Conclusions \& significance}

In this article we have proposed a novel data-driven strategy to dynamically assess the utility of a turbulence modeling hypothesis in an LES framework. This strategy is built on the hypothesis that DNS data may be utilized to assess areas where structural or functional models may be more appropriate in an LES deployment. Our hypothesis segregation and subsequent training culiminates in a learning that may deployed as a classifier of turbulence models at each point on the LES grid as well as a blending technique for balancing turbulence models with different dissipative strengths. When deployed as a classifier, our proposed framework may also predict a `no-model' situation wherein no sub-grid source-term is deployed. When deployed as a blending mechanism, the learning linearly combines the AD and static Smagorinsky hypothesis premultiplied by their respective conditional probabilities to obtain another hybrid dissipation mechanism. Both frameworks utilize the same learning and are assessed through similar experiments in \emph{a posteriori}.

We have rigorously assessed the deployment of our machine learning strategy through the utilization of a Kraichnan turbulence test case. Our assessments are made for Reynolds number values both within and outside that utilized in training to ensure that a generalizable turbulence closure has been developed. In addition, we have also assessed if the proposed closure can be deployed on a coarser grid than one it was trained for. The dissipative and scale-content capture of the proposed framework is compared to the AD and DS techniques through the use of kinetic-energy spectra, vorticity structure functions and time-histories of TKE and vorticity-variance showing a dynamic dissipation akin to the DS. In particular, the statistical fidelity of the data-driven frameworks is seen to be inferior to the AD technique, which provides better estimates of the kinetic energy spectra at lower wavenumbers and also provides most accurate estimates of the vorticity structure function. However, the focus on high-wavenumber noise attentuation leads to no grid cut-off error accummulation and the statistical results of the ML models are very similar to DS in all assessments. 
Also, it is observed that the data-driven closure (whether deployed as a classifier or a blending instrument) adequately captures the $k^{-3}$ scaling expected for the kinetic energy spectra for the Kraichnan turbulence case and attempts to strike an optimal balance between the dissipative functional kernel and the noise-prone structural kernel. This behavior is interesting as the model classifies solely between AD and the static Smagorinsky hypothesis indicating the extreme dissipation of the latter at $C_s=1.0$ is effectively alleviated by the spatiotemporal blending. Our closure, thus, attempts to blend the strengths of both modeling strategies to overcome their individual weakness while attempting to preserve trends from DNS. 

In terms of future opportunities for this idea, the data-driven element of closure identification lends to the potential development of closures that may discern the physical characteristics of different flow scenarios. However, some challenges associated with progress in this research include considerations of invariance properties, which we have identified as a next step. While the computational costs of the proposed framework have not been studied in detail, an efficient deployment of the proposed framework would need graphical processing unit integration of any practical CFD simulation. The latter would lead to efficient learning queries since all the spatial domain information would be available to the common memory. Another future direction identified in this research is the exposure of different two-dimensional turbulence physics to the classification framework to identify if closure choices can ralso be influenced by the training data regime. Success in that regard would allow for `train and forget' closures in problems that have unsteady physics that span fundamentally different turbulence modeling requirements.

\section*{Acknowledgement}

This material is based upon work supported by the U.S. Department of Energy, Office of Science, Office of Advanced Scientific Computing Research under Award Number DE-SC0019290. Disclaimer: This report was prepared as an account of work sponsored by an agency of the United States Government. Neither the United States Government nor any agency thereof, nor any of their employees, makes any warranty, express or implied, or assumes any legal liability or responsibility for the accuracy, completeness, or usefulness of any information, apparatus, product, or process disclosed, or represents that its use would not infringe privately owned rights. Reference herein to any specific commercial product, process, or service by trade name, trademark, manufacturer, or otherwise does not necessarily constitute or imply its endorsement, recommendation, or favoring by the United States Government or any agency thereof. The views and opinions of authors expressed herein do not necessarily state or reflect those of the United States Government or any agency thereof.

\bibliography{jfm-references}
\bibliographystyle{jfm}

\end{document}